\documentclass[preprint2]{aastex}
\usepackage{amsmath}

\shorttitle{A new exact spherical accretion solution}
\shortauthors{X. Hernandez, L. Nasser and A. Aguayo-Ortiz}

\begin{document}


\title{A new hydrodynamic spherical accretion exact solution and its quasi-spherical perturbations}


\author{X. Hernandez$^{1}$, L. Nasser$^{2}$ and A. Aguayo-Ortiz$^{1}$}
\affil{$^{1}$Instituto de Astronom\'{\i}a, Universidad Nacional Aut\'{o}noma de M\'{e}xico, Apartado Postal 70--264 C.P. 
04510 M\'exico D.F. M\'exico.\\
$^{2}$ Department of Science and Mathematics, Columbia College, Chicago, IL 60605, USA. \\}



\begin{abstract}

  We present an exact $\gamma=5/3$ spherical accretion solution which modifies the Bondi boundary condition of $\rho \to const.$ as
  $r\to \infty$ to $\rho \to 0$ as $r \to \infty$. This change allows for simple power law solutions on the density
  and infall velocity fields, ranging from a cold empty free-fall condition where pressure tends to zero, to a hot hydrostatic equilibrium limit
  with no infall velocity. As in the case of the Bondi solution, a maximum accretion rate appears. As in the $\gamma=5/3$ case of the Bondi
  solution, no sonic radius appears, this time however, because the flow is always characterised by a constant Mach number. This number equals
  1 for the case of the maximum accretion rate, diverges towards the cold empty state, and becomes subsonic towards the hydrostatic
  equilibrium limit. It can be shown that in the limit as {  $r \to 0$}, the Bondi solution tends to the new solution presented,
  {  extending the validity of the Bondi accretion value to} cases where the accretion density profile does not remain at
  a fixed constant value out to infinity. We then explore small deviations from sphericity and the presence of angular momentum
  through an analytic perturbative analysis. Such perturbed solutions yield a rich phenomenology through density and velocity fields
  in terms of Legendre polynomials, which we begin to explore for simple angular velocity boundary conditions having zeros on the plane and
  pole. The new solution presented provides complementary physical insight into accretion problems in general.
  
\end{abstract}



\section{Introduction}

It is well established that the infall of gas in the central gravitational potential of accreting objects is the mechanism powering a vast range
of astrophysical phenomena across many orders of magnitude in mass and scale, e.g. Hawley et al. (2015). From the jets originating
about young stellar objects (YSOs), the powerful emissions of gamma ray bursts (GRBs) to the active galactic nuclei harbouring accreting
super massive black holes (SMBHs) in their centers, it is accretion that ultimately powers the observed outflows.

The study of the hydrodynamics of astrophysical accretion processes started with the seminal work of Bondi (1952), where spherical symmetry
is assumed about a point mass, together with the boundary condition of a density which tends to a constant value for large radii. This model,
together with its relativistic extension by Michel (1972), has provided valuable insight into the magnitude of accretion rates onto central
objects and the physical scalings of this process with the mass of the accretor and the densities and temperatures of the accreting material.
{  The above in spite of the lack of {\bf explicit} analytical expressions for the infall velocity and the density fields in the Bondi Model.}


In this paper we modify the Bondi boundary conditions
to a density profile which tends to zero at infinity, which for a $\gamma=5/3$ equation of state and a Keplerian potential allows for a simple
analytic power law steady-state solution. Hence, this solution might be of particular relevance to systems where accretion density
profiles do not quickly converge to constant values outwards of the Bondi radius, but exhibit falling density profiles for large dynamical ranges.

In section 2 we present a new exact analytic spherically symmetric accretion model for $\gamma=5/3$, explore some of its scalings and
implications and present a comparison to the corresponding Bondi model {  and to a couple of observed accretion density profiles in
nearby x-ray emitting elliptical galaxies from the sample of Pl\v{s}ek et al. (2022)}. Our new solution is then used in section 3 as the basis
for a first-order perturbative analysis preserving axial symmetry, but to first order on departures from sphericity with the polar angle and
considering the inclusion of angular momentum, also to first order. Then, section 4 briefly presents two particular examples of the latter,
{  and section 5 a first numerical exploration of the new spherical solution obtained and of its simplest polar angle deviations from sphericity.
Section 6} presents our conclusions. Appendix A shows fits of our new spherical accretion model to the full sample of observed accretion density
profiles from Pl\v{s}ek et al. (2022), {  and finally, appendix B} presents a perturbative analysis for deviations from
$\gamma=5/3$ for the spherically-symmetric model.

\section{Spherical $\gamma=5/3$ accretion model}

We begin with the steady-state equations of conservation of mass and radial and angular momentum for a gas distribution in the
presence of a Newtonian gravitational potential produced by a point mass $M$. Assuming axial symmetry and using a spherical
coordinate system with $\theta$ the angle between the positive vertical direction and the position vector $\vec{r}$ {  and $\phi$
an azimuthal angle} we have:

\begin{equation}
\frac{1}{r^{2}} \frac{\partial(r^{2} \rho V)}{\partial r} =- \frac{1}{r \sin \theta}\frac{\partial(\rho U \sin \theta)}
{\partial \theta},
\end{equation}

\begin{equation}
V\frac{\partial V}{\partial r} +\frac{U}{r}\frac{\partial V}{\partial \theta} -\frac{U^{2}}{r} -\frac{W^{2}}{r}=
-\frac{1}{\rho}\frac{\partial P}{\partial r} - \frac{G M}{r^{2}},
\end{equation}

\begin{equation}
  V\frac{\partial U}{\partial r} +\frac{U}{r}\frac{\partial U}{\partial \theta} +\frac{V U}{r}-\frac{W^{2} \cot \theta}{r}
  =-\frac{1}{r \rho}\frac{\partial P}{\partial \theta},
\end{equation}

\begin{equation}
  V\frac{\partial W}{\partial r} +\frac{U}{r}\frac{\partial W}{\partial \theta} +\frac{VW}{r} +\frac{UW \cot \theta}{r}=0
\end{equation}

\noindent where $\rho(r,\theta)$, $P(r,\theta)$, $V(r,\theta)$, $U(r,\theta)$ and $W(r, \theta)$ are the gas density and pressure, and the $r$, $\theta$
and $\phi$ velocities, respectively e.g. Binney \& Tremaine (1987). Assuming a barotropic equation of state $P=K \rho^{\gamma}$ eqs.(2) and (3) become:

\begin{equation}
V\frac{\partial V}{\partial r} +\frac{U}{r}\frac{\partial V}{\partial \theta} -\frac{U^{2}}{r} -\frac{W^{2}}{r} =
-K \gamma \rho^{\gamma-2} \frac{\partial \rho}{\partial r} - \frac{G M}{r^{2}},
\end{equation}

\begin{equation}
  V\frac{\partial U}{\partial r} +\frac{U}{r}\frac{\partial U}{\partial \theta} +\frac{V U}{r} -\frac{W^{2} \cot \theta}{r}
  = -\frac{K \gamma \rho^{\gamma-2}}{r}  \frac{\partial \rho}{\partial \theta},
\end{equation}

We now write the above equations in dimensionless form introducing the variables $\varrho=\rho/\bar{\rho}$,
$\mathcal{V}=V/\bar{c}$, $\mathcal{U}=U/\bar{c}$, $\mathcal{W}=W/\bar{c}$ and $R=r/\bar{r}$, where $\bar{\rho}$ is a reference density at a certain point,
$\bar{c}^{2}=K \gamma \bar{\rho}^{\gamma-1}$, the sound speed at this same reference point, and $\bar{r}=GM/\bar{c}^{2}$. Equations (1), (5), (6)
and (4) now read:

\begin{equation}
\frac{\partial(R^{2} \varrho \mathcal{V})}{\partial R} =- \frac{R}{\sin \theta}\frac{\partial(\sin \theta  \varrho \mathcal{U})}
{\partial \theta},
\end{equation}

\begin{equation}
  R\mathcal{V}\frac{\partial \mathcal{V}}{\partial R} +\mathcal{U}\frac{\partial \mathcal{V}}{\partial \theta}
  -\mathcal{U}^{2} -\mathcal{W}^{2} = -R\varrho^{\gamma-2} \frac{\partial \varrho}{\partial R} - \frac{1}{R},
\end{equation}

\begin{equation}
 R \mathcal{V}\frac{\partial \mathcal{U}}{\partial R} +\mathcal{U}\frac{\partial \mathcal{U}}{\partial \theta}
  +\mathcal{V} \mathcal{U} -\mathcal{W}^{2} \cot\theta =-\varrho^{\gamma-2}\frac{\partial \varrho}{\partial \theta}.
\end{equation}

\begin{equation}
  R \mathcal{V} \frac{\partial \mathcal{W}}{\partial R} +\mathcal{U}\frac{\partial \mathcal{W}}{\partial \theta}
  +\mathcal{V}\mathcal{W}+\mathcal{U}\mathcal{W} \cot\theta=0.
\end{equation}

\noindent Equations (7)-(10) now define the general problem. The first step is to obtain a solution for the unperturbed state, which
will be one of spherical accretion {  and zero angular momentum, $(\partial \varrho/\partial \theta)= \mathcal{U}= \mathcal{W}=0$,}
and hence a solution to:

\begin{equation}
\frac{d(R^{2} \varrho \mathcal{V})}{d R} =0,
\end{equation}

\begin{equation}
  R\mathcal{V}\frac{d \mathcal{V}}{d R} = -R\varrho^{\gamma-2} \frac{d \varrho}{d R} - \frac{1}{R},
\end{equation}

\noindent where at this point we have assumed $\mathcal{\varrho}$ and $\mathcal{V}$ are both functions of $R$ alone.
Although this problem is generally treated in terms of the Bondi (1952) solution, we change the Bondi boundary condition of $\varrho=const.$ as $R \to \infty$
for $\varrho=0$ as $R \to \infty$. Whilst the assumptions of strict spherical symmetry, zero angular momentum and an adiabatic index of
exactly $\gamma= 5/3$, will never be realised in any actual astrophysical setting, we introduce them here to obtain an exact solution,
which can then be used to study perturbatively modifications introduced by the inclusion of small amounts of angular momentum and angular
dependences in velocity and density fields, as well as small variations about $\gamma =5/3$.

{  We seek a new solution to eqs. (11) and (12) by imposing an assumption of radial power laws for the velocity and density fields,
  and then try to find such functions guided by the structure of these two equations. The last term of eq.(12), the Newtonian point
  mass potential, fixes that the other two terms of this same equation must necessarily also scale with $R^{-1}$. For the case of
  the velocity term, the scaling imposed by the potential uniquely determines the velocity power-law scaling as $\mathcal{V}(R) \propto R^{-1/2}$.
  Going back to the mass conservation equation, eq. (11), the above velocity scaling now uniquely fixes the density scaling as $\varrho(R) \propto R^{-3/2}$.
  Now, the second term in eq.(12) will also be a power-law, with a scaling of $R^{3(1-\gamma)/2}$. This last will only correspond to the
  $R^{-1}$ scaling imposed by the Newtonian potential in eq.(12) precisely and uniquely for the value of $\gamma=5/3$ we are interested in. The velocity and
  density scalings found above are both negative and hence we see that they satisfy our required boundary conditions.

  Thus, we see that imposing
  only spherical symmetry, steady state accretion, a Newtonian point mass potential and power law solutions for both the density and velocity fields,
  uniquely determines $\varrho(R) \propto R^{-3/2}$, $\mathcal{V}(R) \propto R^{-1/2}$ and $\gamma=5/3$. No other power law
  solution to the steady state spherically symmetric accretion problem exists for a Keplerian potential. Under this scheme, $\gamma=5/3$ becomes a constraint,
  which happens to correspond to a value of the adiabatic index often considered in a variety of astrophysical scenarios.

  Our new solution is hence clearly
  less general than the Bondi solution with respect to the thermodynamic range over which it is valid; while a broad range of adiabatic indices are consistent
  with the Bondi framework, the solution presented here is only valid for $\gamma=5/3$, although perturbative deviations from this value can be
  formally treated, see appendix B. The ansatz of power law solutions
  introduced follows from analytic simplicity considerations. However, the particular unique power law scalings found are determined by
  the physics of the assumptions introduced, having taken a stationary adiabatic spherically symmetric accretion and a keplerian potential.}

Therefore equations (11) and (12) allow a closed analytic solution for the natural choice of $\gamma=5/3$, using only radial power laws described by:

\begin{equation}
\mathcal{V}=\mathcal{V}_{0}R^{-1/2},
\end{equation}

\begin{equation}
\varrho=\varrho_{0}R^{-3/2},  
\end{equation}

\noindent where $\mathcal{V}_{0}$ and $\varrho_{0}$ are two constants satisfying eq.(12):

\begin{equation}
\mathcal{V}_{0}^{2}=2-3\varrho_{0}^{2/3}.  
\end{equation}

\noindent Thus, the full boundary conditions at infinity are for the fluid at rest with zero density
and zero sound speed. This fixes the Bernoulli equation with constant $=0$, as in the Bondi solution.
It is clear that eq.(11) can be integrated directly to yield the dimensionless mass accretion rate as:

\begin{equation}
\dot{M}=4 \pi \mathcal{V}_{0}\varrho_{0}.
\end{equation}

Equation (4) in Bondi (1952) is the Bernoulli equation for the problem, a first integral of the momentum equation
solved for non-zero pressure and density at infinity. Here we do not make use of the Bernoulli relation, nor impose the existence
of a sonic radius {  (or its non-existence {\it a priori} either)}, but rather solve the conservation equations directly.
Therefore, the solution of eqs. (13) and (14) does not lie within the Bondi framework and constitutes a new solution. \footnote{When going
through Bondi (1952) one must be cautious of the typos in eq.(12) and eq.(13) which should read : $y=u^{2/(\gamma+1)}(\lambda/x^{2})^{(\gamma-1)/(\gamma+1)}$
and $z=(\lambda/x^{2}u)^{2/(\gamma+1)}$, respectively.}

\begin{figure}
\hskip -10pt \includegraphics[width=9.25cm,height=8.0cm]{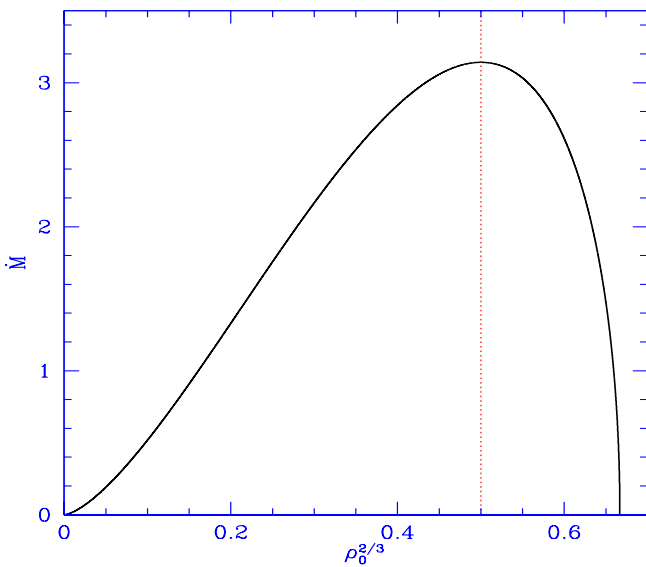}
\caption{The figure shows the dimensionless accretion rate, $\dot{M}$, as a function of the dimensionless density
  parameter, $\varrho_{0}^{2/3}$, for the spherically symmetric solution. $\dot{M}$ goes to zero both for the cold, empty free-fall
  limit at $\varrho_{0}^{2/3}=0$ and for the hot, hydrostatic equilibrium one at  $\varrho_{0}^{2/3}=2/3$, with a maximum accretion
  rate of $\dot{M}_{c}=\pi$ at $\varrho_{0}^{2/3}=\mathcal{V}_{0}^{2}=1/2$.}
\end{figure}

\begin{figure}
\hskip -10pt \includegraphics[width=9.25cm,height=8.0cm]{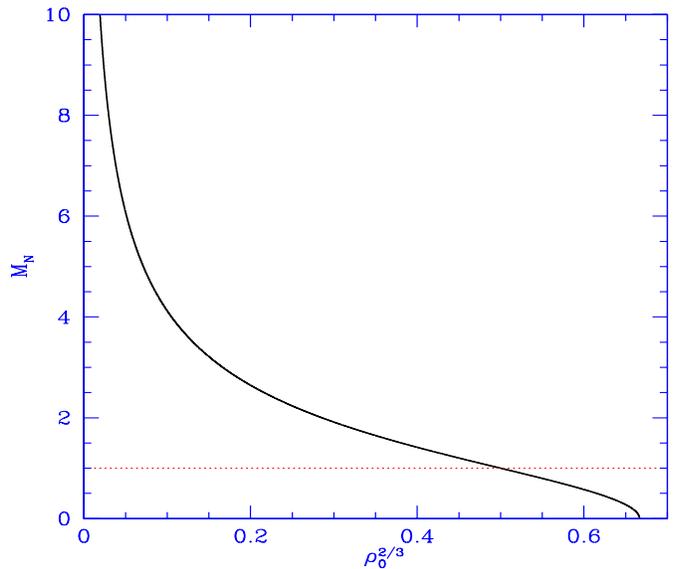}
\caption{For the spherically symmetric solution, the solid curve gives the Mach number, which is a constant with radius, as a function of the
  dimensionless density parameter, $\varrho_{0}^{2/3}$. This diverges for the cold, empty free-fall $\varrho_{0}^{2/3}=0$ limit, and goes to
  zero at the hydrostatic equilibrium $\varrho_{0}^{2/3}=2/3$ one. The dotted line gives $\mathcal{M}=1$, which is crossed at
  $\varrho_{0}^{2/3}=\mathcal{V}_{0}^{2}=1/2$, corresponding to the maximum dimensionless accretion rate $\dot{M}_{c}=\pi$.}
\end{figure}

We see from eq.(15) that the solution will only exist for the interval $0<\varrho_{0}^{2/3}<2/3$, over which the
$\mathcal{V}_{0}$ constant determining the amplitude of the inward velocity flow, and which hence implies that this
constant is negative, will vary within the range $2>\mathcal{V}_{0}^{2}>0$. As $\varrho_{0} \to 0$ pressure tends to zero and we
reach an empty state in Keplerian free-fall. As $\varrho_{0}^{2/3} \to 2/3$ we approach a dense hydrostatic equilibrium configuration where
the infall velocity tends to zero. The accretion rate is clearly zero at these two limits, and has a maximum at some intermediary
value of $\varrho_{0}^{2/3}$, which is hence seen as the sole parameter of the spherical solution. We can use eq.(16) to write eq.(15)
in terms of $\varrho_{0}^{2/3}$ and $\dot{M}$ only:

\begin{figure*}
  \hskip -20pt \includegraphics[height=8cm,width=8.5cm]{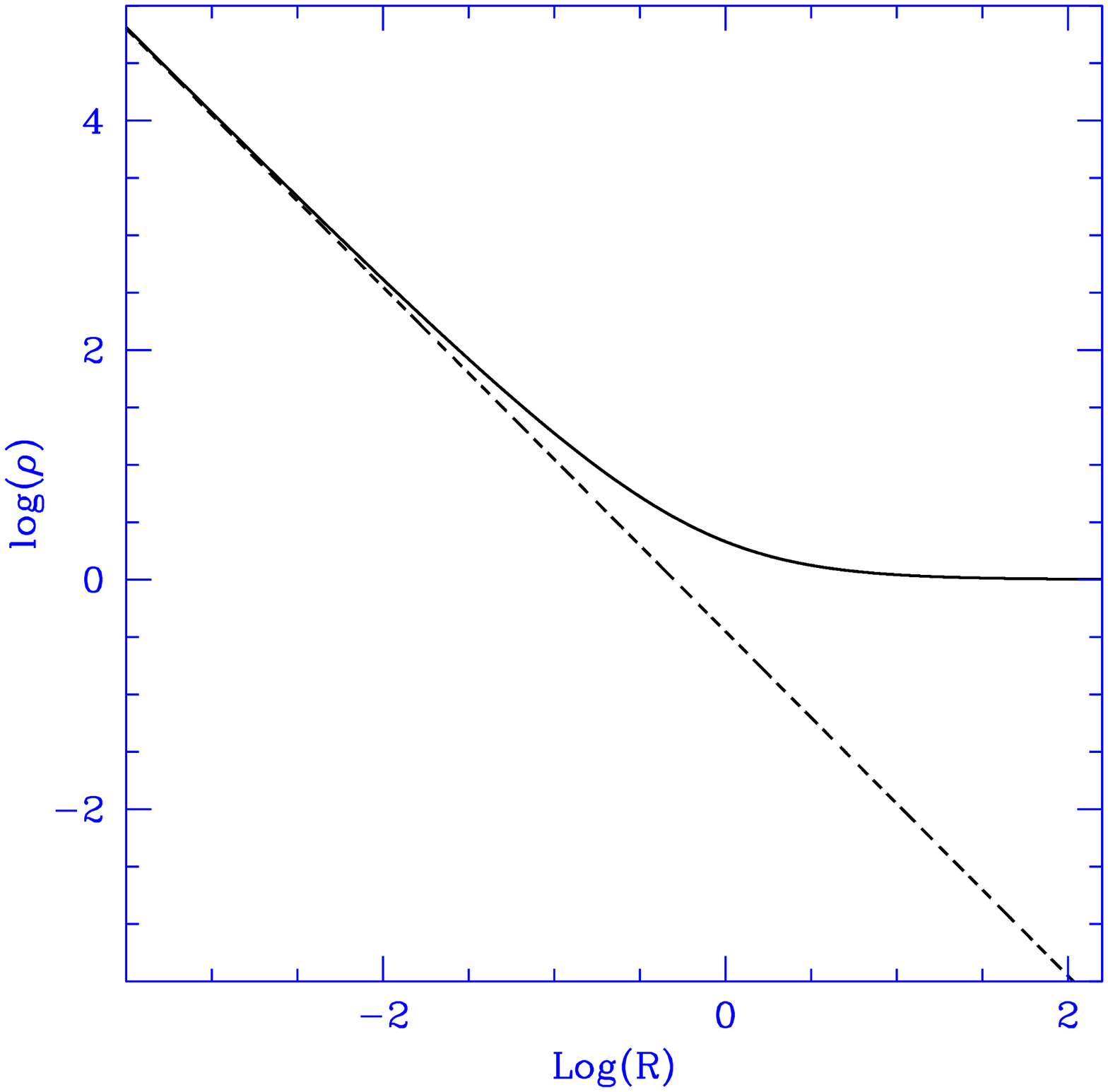}
  \includegraphics[height=8cm,  width=8.5cm ]{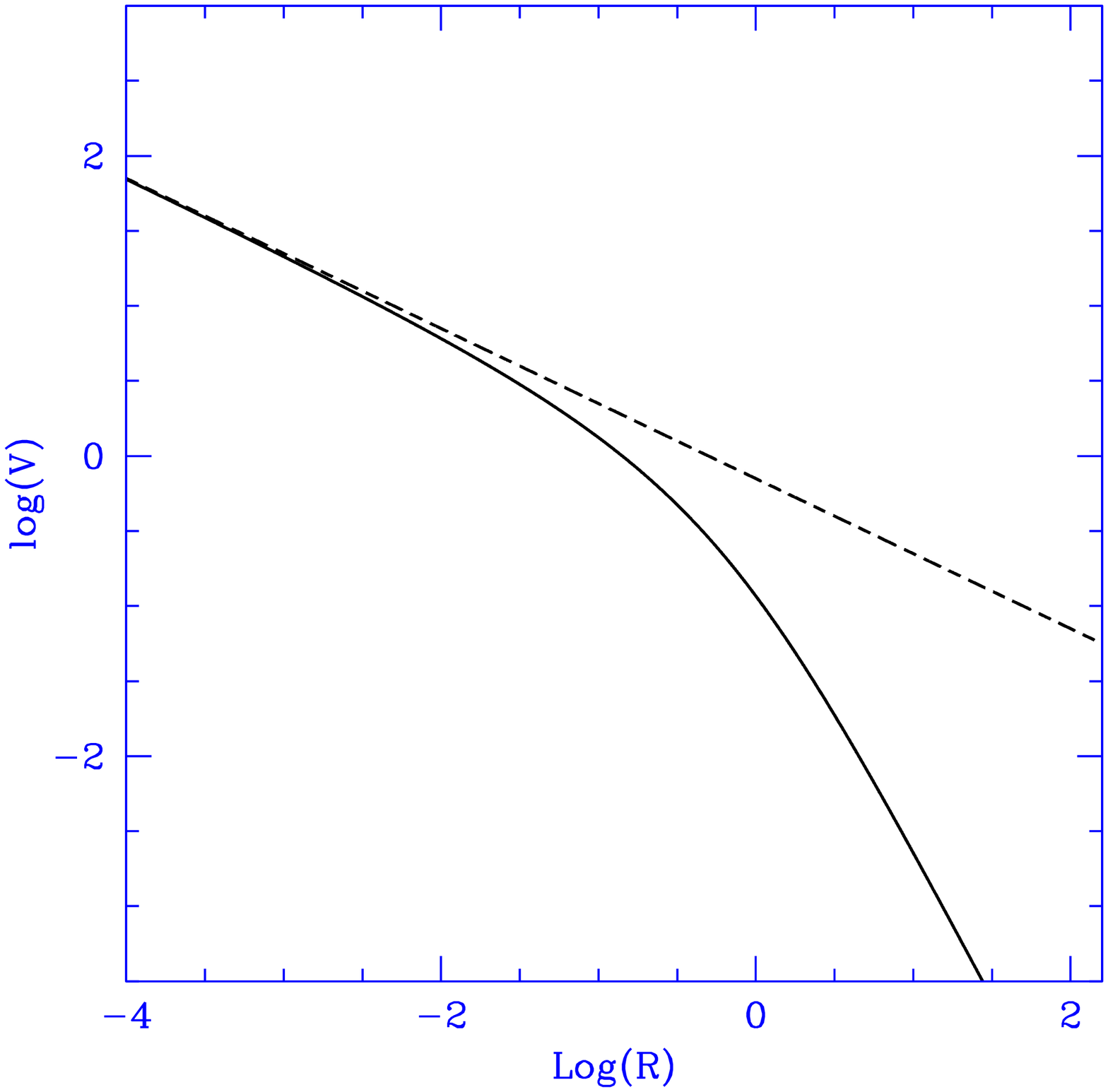} 
  \caption{  {Left}: The solid curve gives the $\gamma=5/3$ density profile of the Bondi solution, clearly showing the large {  radii} constant
    density behaviour, {  strongly contrasting with the decaying profile of our new solution}. The dashed line gives the density profile of eq.(14)
    for $\mathcal{M}=1$, clearly a profile to which the Bondi solution
    converges for small radii. {Right}: The solid curve gives the $\gamma=5/3$ velocity profile of the Bondi solution, while the dashed line
    gives the velocity profile of eq.(13) for $\mathcal{M}=1$, clearly a profile to which the Bondi solution converges for small radii,
    {  whilst again a clear divergence between the two is evident at large radii. In both cases the radial scale is given in units of the
    Bondi radius.}}
\end{figure*}

\begin{equation}
\dot{M}^{2}=(4 \pi \varrho_{0}^{2})^{2}\left(2- 3\varrho_{0}^{2/3}  \right).
\end{equation}

The above equation can now be differentiated w.r.t. $\varrho_{0}$ to obtain the maximum value of the accretion rate. This will occur at
$\varrho_{0c}^{2/3}=\mathcal{V}_{0c}^{2}=1/2$ at a value of $\dot{M}_{c}=\pi$. Fig. 1 gives a plot of $\dot{M}$ as a function of $\varrho_{0}$
in the range of interest. In physical units we obtain:

\begin{equation}
\dot{M}_{c,ph}=\pi \frac{\bar{\rho}}{\bar{c}^{3}}(GM)^{2},
\end{equation}

\noindent which can be written in terms of the barotropic constant of the problem $K$ as:

\begin{equation}
\dot{M}_{c,ph}=\pi(GM)^{2} \left( \frac{3}{5K} \right)^{3/2}.
\end{equation}

\noindent Thus, we see that in terms of physical units, the problem is fully determined by the constant $K$ which can be fixed by
specifying a pair of density and sound speed values at any given radius.

{  Notice that the framing of the solution of eqs. (13) and (14) in the context of accretion determines the sign of the $\mathcal{V}_{0}$
constant as negative. However, as can be seen from eq. (15), this sign could just as well be positive and the resulting ejection
solution will still solve exactly the conservation equations under the assumptions of spherical symmetry and  $\gamma=5/3$. Such a solution
could represent the asymptotic consequence of a wind expanding in a vacuum and clearly also deserves attention and further study. Still,
in this first exploration of the problem we choose to concentrate on accretion scenarios and defer study of the ejection branch of the
solution, its temporal, thermodynamic and non-spherical perturbations, to future works.}

We complete the study of the spherically symmetric adiabatic accretion solution with an evaluation of the Mach number of the flow,
which can be derived by writing $\mathcal{V}$ from eq. (13) in physical units by multiplying by $\bar{c}$, and dividing by
the local physical sound speed $c^{2}=(5K/3)\rho^{2/3}$, with $\rho$ written as $\varrho(R) \times \bar{\rho}$ we obtain:

\begin{equation}
\mathcal{M}=\mathcal{V}_{0}\varrho_{0}^{-1/3}.
\end{equation}

We see that for the maximum accretion rate at $\varrho_{0}^{2/3}=\mathcal{V}_{0}^{2}=1/2$, the Mach number
of the flow is precisely $\mathcal{M}=1$. In going towards the hotter and denser subsonic configurations towards the $\varrho_{0}=2/3$
of hydrostatic equilibrium, the Mach number drops gradually to zero, while for less dense supersonic cold models with
$\varrho_{0}^{2/3}<1/2$, the Mach number gradually diverges on approaching $\varrho_{0}^{2/3}=0$ as the sound speed goes to zero.
A plot of this is given in fig. (2). Whilst in the Bondi solution a sonic radius exists for $\gamma \neq 5/3$, and the monoatomic
adiabatic value is a singular point where the sonic radius goes to zero, in the power law solutions presented here there is no sonic
radius as $\mathcal{M}$ becomes a constant with radius, with all sub-maximum accretion models being either supersonic or subsonic
at all radii, depending on whether the value of $\varrho_{0}$ is chosen above or below the critical one corresponding to maximal
accretion at $\mathcal{M}=1$. {  Hence, for the new solution presented the flow is never trans-sonic and no shocks will develop.}

The spherical accretion solution of eqs. (13) and
(14) could arise from the inward propagation of a rarefaction wave into gas initially at rest with a density profile decreasing to zero at
infinity. This has to be explored through both analytical and numerical dynamical studies so as to asses the initial condition parameter space in
both the velocity and density fields to constrain which regions might result in the $\gamma=5/3$ constant Mach number solution presented,
as a consequence of the dynamics of the propagation of the rarefaction wave, the details of the initial conditions, or a combination of both.
The above lies beyond the scope of the first presentation of the new spherical accretion presented here and will be treated in subsequent studies.

That the solution presented here does not form part of the Bondi framework becomes evident since we have explicitly shown exact non-transonic
solutions to the spherically symmetric accretion problem with radially constant values of $\mathcal{M}$. 
As the local sound speed is given by $(dP/d \rho)^{1/2}$, and given the barotropic equation of state used of $P \propto \rho^{\gamma}$,
the sound speed will scale as $c^{2} \propto \rho^{\gamma-1}$, for $\gamma=5/3$ we get $c \propto \rho^{1/3}$. For the exact solution of
the conservation equations presented in eqs. (13) and (14), as $\rho \propto R^{-3/2}$, we get $c \propto \rho^{-1/2}$, exactly
the same radial scaling of the velocity flow in eq.(13), forcing a radially constant Mach number. Thus, the mach number of the flow is a
constant with radius, at a value fixed once $\varrho_{0}$ is chosen, which in turn determines $\mathcal{V}_{0}$ through eq.(15).

{  Notice that we have in no way either imposed at the onset or forced the lack of a sonic radius at any point in the development leading to
the new solution found. This unexpected feature of our solution follows from the new exact power law solutions to both the mass conservation and
radial Euler equations we found. We do not know if this particular feature holds only for the solution we have found, or if there might exist other
more complex non-power law solutions where also no sonic radius exists and the Mach number appears a radial constant. It is entirely possible that
just as we have found a simple exact spherical accretion solution where no sonic radius appears and $\mathcal{M}$ is constant, there might exist more
complex non-power law solutions where this feature is maintained. However}, as shown and discussed in appendix B, even perturbative departures from
$\gamma=5/3$ result in unequal changes to the local sound speed and the flow velocity scalings, implying the presence of a sonic radius for even the
slightest deviation from  $\gamma=5/3$.

We can now compare our $\mathcal{M}=1$ maximum accretion rate solution explicitly to the maximum accretion rate of the Bondi solution
(usually referred to simply as the Bondi solution) for $\gamma =5/3$, where the density field satisfies the following implicit relation:

\begin{equation}
(1/2)^{4}+3R^{4}\varrho_{B}^{8/3}=3 R^{4}\varrho_{B}^{2}+2R^{3}\varrho_{B}^{2},  
\end{equation}

\noindent which is equation (15) in Bondi (1952) for $\gamma=5/3$, where $\varrho_{B}$ is the dimensionless density in units of the density at
infinity assumed by Bondi. Notice that substituting a $R^{-3/2}$ density scaling in the above equation will result in a divergence in the first
term in the R.H.S., showing once again that the solution of eqs.(13)-(15) lies outside of the Bondi framework.

It is clear from eq.(21) that as $R \to 0$, the first term in the right hand side can be dropped, as the second term in the
right hand side will dominate. The three remaining terms must hence balance, leading to $\varrho_{B} \propto R^{-3/2}$. Mass conservation
through eq.(11) now implies $\mathcal{V}_{B} \propto R^{-1/2}$, fulfilling the match to the scalings of eqs.(13) and (14).

Introducing the mass conservation equation for the Bondi solution for $\gamma=5/3$ of $R^{2}\varrho_{B} \mathcal{V}_{B}=1/4$, allows to evaluate the
scaling constants of the small radius asymptote of the Bondi solution as $(1/2)^{3/2}$ and $(1/2)^{1/2}$ for the density and velocity field respectively,
exactly the values of our new solution. Hence, we see that the small radius limit of the $\gamma =5/3$ Bondi solution exactly corresponds to the
solution we have presented. 

The full comparison of the density and velocity fields of the $\gamma=5/3$ Bondi solution to our new one is presented in fig. 3, showing clearly
the convergence proven above for small radii. It is also evident from fig 3, unavoidable since the boundary conditions at infinity for our
solution and the Bondi case are very different, that both solutions strongly diverge as the radius grows larger than the Bondi radius.
Given that the accretion rate is a constant with radius, we see that the well established scalings of the Bondi accretion rate with density and sound
speed, do not require the formal validity of the Bondi density field, which remains at a finite value out to infinity, and are in fact realised also
for the solution presented, where the density falls to zero as $R^{-3/2}$.

\begin{figure*}
  \includegraphics[height=8cm,  width=8.5cm  ]{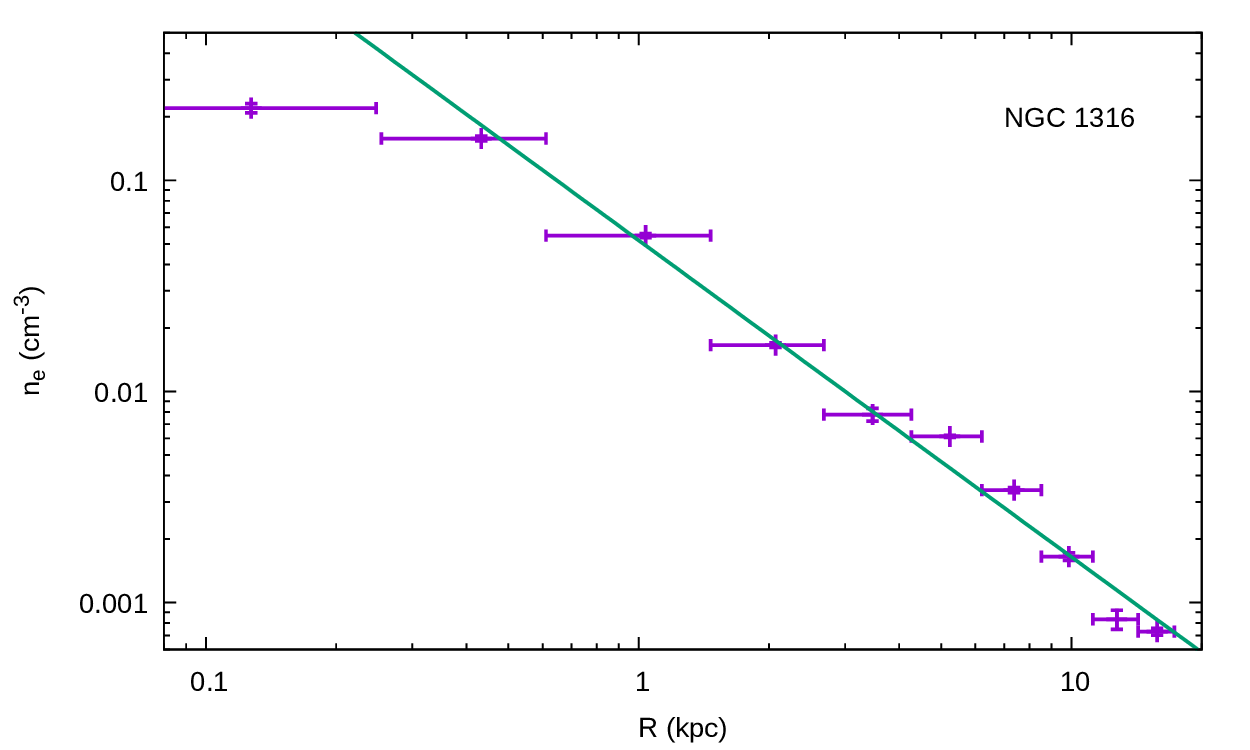} \includegraphics[height=8cm,  width=8.5cm  ]{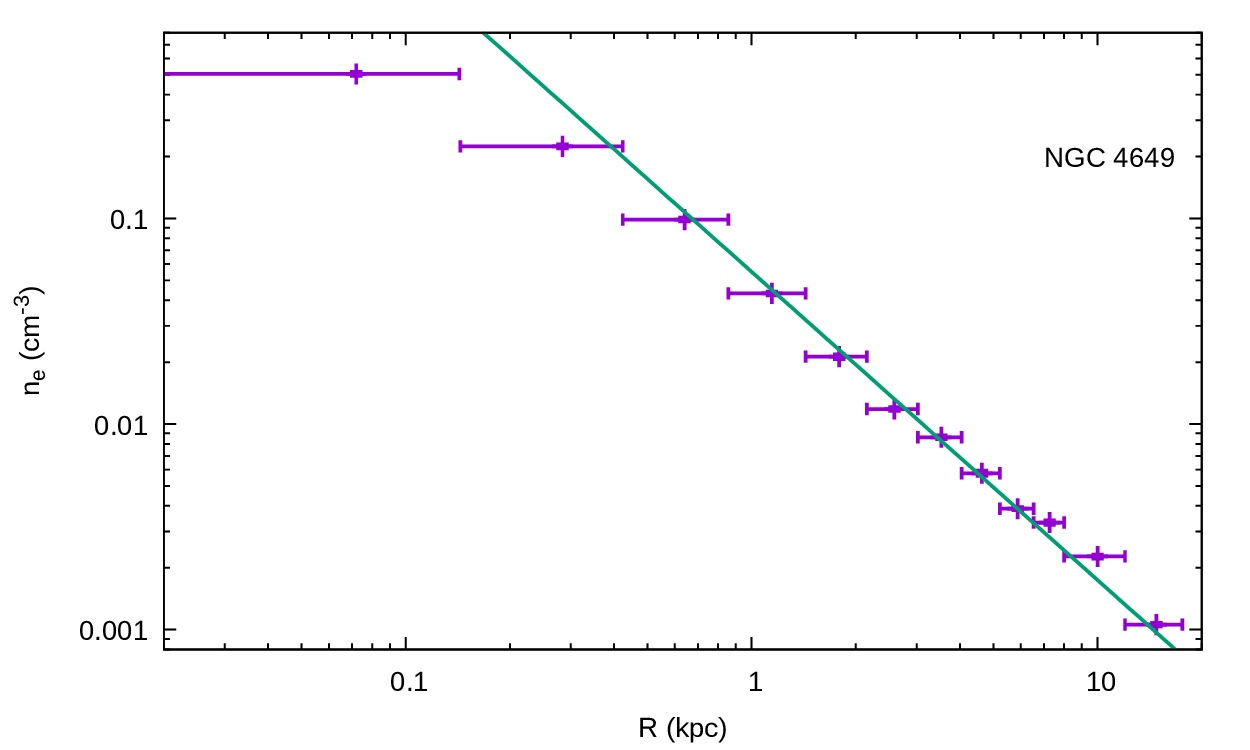} 
  \caption{Inferred electron density profiles about the central SMBHs of two galaxies as presented by Pl\v{s}ek et al. (2022), points with error bars.
    The solid lines give $R^{-3/2}$ fits, as expected by the new solution of eqs. (13) and (14). The Bondi radii of the two galaxies are of
    $7.4^{+1.3}_{-1.3}$ pc and $111 \pm 19$ pc for NGC 1316 and NGC 4649, respectively, as reported by Pl\v{s}ek et al. (2022). Classical Bondi fits
    would look very close to horizontal over the radial range shown, given the constraint of the inferred values for the Bondi radii.
  }
\end{figure*}

Notice that the convergence proven for the maximum accretion classical Bondi solution at $\gamma=5/3$ to the maximum accretion rate of our new
$\gamma =5/3$ solution, applies only for the limit as $R\to 0$. Indeed, it is crucial to appreciate that the Bondi solution implies
more than just a scaling between a reference density and sound speed and a resulting accretion rate; it implies also particular functional forms
for the density and velocity profiles of the accreting material. In particular, the density profile of the Bondi solutions is such that a constant
asymptotic value is rapidly approached outwards of the reference Bondi radius. This might be a reasonable description of some actual astrophysical
situations, but not of others. For example, the observed accretion density profiles in X-ray luminous elliptical galaxies reported by Allen et. al
(2006), show no convergence towards finite values out to many tens and in some cases even hundreds of Bondi radii. In fact, the above mentioned
accretion density profiles are clearly falling consistently as the radius increases, out to the last observed point, much more in line with the
new solution presented here where the accretion density profile tends to zero at large radii, than with the expectations of the classical
Bondi models.

In terms of a quantitative comparison of the new spherically symmetric accretion solution presented here and observations, the works of
Wong et al. (2014), Russell et al. (2015) and Runge \& Walker (2021) inferring deprojected hot X-ray accretion density profiles in NGC 3115, M81 and
NGC 1600, respectively, are relevant. M81 and NGC 1600 show substantial AGN activity tending to flatten the observed accretion density profiles
outside of the Bondi radius (Runge \& Walker 2021), which have power-law slopes of $-0.41 \pm 0.04$ and $-0.65 \pm 0.04 $ respectively, for radial
ranges larger than the Bondi radii. Again, accretion density profiles which do not tend to a constant out to many tens of their Bondi radii, with
the expectation that in the absence of AGN activity, these two density profiles would be steeper. Finally, NGC 3115 shows a decreasing power-law
density profile beyond the Bondi radius with a slope of $-1.34^{+0.25}_{-0.2}$, consistent with the expectations of the new model presented here of
$\varrho \propto R^{-3/2}$. Interestingly, this last galaxy is a quiescent system with no AGN activity to perturb the accretion density profile.

Similarly, accretion density profiles for hot x-ray emitting gas using {\it Chandra} observations about SMBHs for a large sample of 20 nearby galaxies
are reported by Pl\v{s}ek et al. (2022), all of which show clearly falling radial profiles beyond the Bondi radius, with power law slopes close to the
expectations of the new model presented here. {  Figure 4 shows the inferred volumetric electron density profiles from Pl\v{s}ek et al. (2022) for NGC
1316 and NGC4649, points with error bars, together with $R^{-3/2}$ fits to the observed profiles. In all cases for the innermost points, the reported
relative radial errors are of 100\%, for which reason the innermost points have been ignored in the fits.

It is clear that the fits are very accurate
representations of the data over the close to two orders of magnitudes in radius covered by the observations. Also, the Bondi radii of the two galaxies
shown in the figure are of $7.4^{+1.2}_{-1.3}$ pc for NGC 1316 and $111 \pm 19$ pc for NGC 4649, as inferred by Pl\v{s}ek et al. (2022). Thus, a Bondi fit
to the observed profiles would appear close to a horizontal line for the case of NGC 1316 over the radial range covered by the data, and beyond 1 kpc a
solution rapidly convergent to a constant value for the case of NGC4649. The rapid flattening of the Bondi profile outwards of the Bondi radius, as seen
in figure (3), shows that once constrained by the inferred values of the Bondi radii, the two cases shown above will yield significantly poorer fits to
classical Bondi profiles than to the expected $R^{-3/2}$ density scalings of the solution in eqs. (13) and (14).

Appendix A presents corresponding fits
to the remaining 18 galaxies in Pl\v{s}ek et al. (2022), showing in most cases excellent agreement to the density radial scaling of our new solution of
$R^{-3/2}$. {  Even} in cases where this scaling is not an excellent fit, it remains a much better representation than the essentially flat expectations
of the classical Bondi solution, for the radial ranges covered. It appears that for hot accretion density profiles about central galactic SMBHs, observationally
inferred profiles are much more consistent with our new solution than with the classical Bondi one. Indeed, we are not aware of even a single astrophysical hot
accretion density profile which is observed to flatten to a constant value outwards of the Bondi radius.}

{  This, however, does not necessarily invalidate
mass accretion rates calculated under the Bondi model for the above observed profiles. The reason being that although the Bondi mass accretion rate is always
proportional to $\varrho_{\infty}/c_{\infty}^{3}$, in the particular case of $\gamma=5/3$ the density at infinity is no longer a critical parameter for the accretion
rate in the Bondi case. Since as $c^{2}=dP/d\varrho$, it is easy to see that the Bondi accretion rate will in general scale with $\varrho^{5/2-3\gamma/2}$,
which for the particular case of $\gamma=5/3$ reduces to a constant given by the entropy of the gas (as happens also in our case c.f. eq.(19)), proportional
to $log (P/\varrho^{\gamma})$. Thus, for $\gamma=5/3$ the density at infinity for the Bondi model ceases to be a critical parameter and the accretion rate
remains constant regardless of the particular value of the density at infinity, provided the gas remains at constant entropy. Hence, in concordance with
the discussion following eq. (21), mass accretion rates generally found in the literature which are calculated using density and sound speed estimates at the
Bondi radius, are consistent with the solution presented here (which converges to the Bondi one for small radii), even if the large-scale accretion density
profiles do not correspond to the detailed Bondi density accretion profile.}

It is clear that the comparison presented above is approximate, in as much as in the physical systems described a number of effects beyond the
simplifying assumptions giving rise to the exact solution of eqs. (13) and (14) are relevant. Some of these can be dealt with through careful
treatment of the observations used, such as the presence of non-thermal components in the x-ray emitting gas arising from low mass x-ray binaries
(LMXB) or AGN jet activity. Indeed, all the references mentioned above include careful spectral analysis to identify only the thermal component of
the x-ray emitting gas when producing de-projected density profiles. Towards the central regions, in some cases thermal instability might set in, leading
to a multi-phase gas structure, as evident through emission line diagnostics. About two thirds of the galaxies in the  Pl\v{s}ek et al. (2022)
sample show such evidence, or the central region of NGC 1600 as studied by Runge \& Walker (2012) where a cold and a hot phase are directly inferred
through spectroscopic fitting to the {\it Chandra} data used. It is reassuring that the data discussed clearly show density profiles consistent with
the solution of eqs.(14) substantially beyond this inner regions. Similarly, other physical processes at play in the actual observed
cases include the non-adiabatic character of the gas flows. Fortunately, as shown by e.g. Sun \& Yang (2021), given the strong density dependences
of both heating (from irradiation by a central x-ray emission) and cooling rates, this lack of adiabaticity will again be confined to the very central
regions, typically at pc scales much smaller than the radial ranges fitted in figure 4 and appendix A.

As with the cases discussed above, many 
other of the more complex physical processes likely at play in a real situation will naturally intensify towards smaller radii, e.g. general
relativistic departures from the Newtonian potential assumed in the solution presented here. It is therefore encouraging that the general good
agreement of the density profile of eq.(14), and indeed its much better agreement with the data than the close to horizontal expectation of a
classical Bondi profile for the Bondi radii inferred in the cases presented, applies to the large scale regions of the observations discussed.

We are aware that the comparisons in density profiles presented here and in the appendix A can not be regarded as any more than approximate, given the
existence of a range of physical processes in the actual cases which have not been included in the idealised model presented. We do not intend the
comparisons presented as proof of the accurate validity of the model presented, although the quite good fits obtained in many cases (see appendix A),
particularly beyond the more problematic central regions, and the fact that these fits are clearly much better than to a similarly approximate
Bondi profile, give us confidence that the solution presented captures some of the physics at play, to a greater degree than what happens
with a classical Bondi profile, which flattens to a constant density on crossing the Bondi radius.
  
Regarding a first quantitative comparison to numerical experiments, Sun \& Yang (2021) model the hydrodynamical accretion of optically transparent hot
gas onto a central black hole for low luminosity AGNs, assuming spherical symmetry and an adiabatic index of $\gamma=5/3$, and obtain density profiles
consistent with the expectations of the new model presented here, see their figure 7. {  Along these lines, in section 5 we present first
numerical simulations of the new spherical solution presented here, showing it to be stable.}

The original idea of Bondi was to model a small star accreting material from a large cloud of constant density material, formally taken as infinite
in size. Thus, the star presents only a small local perturbation on the structure of the cloud, and the accretion density profile rapidly converges
to the asymptotic constant density of the infinite cloud. As can be seen from fig.(3), the Bondi accretion density profile changes qualitatively
on crossing the Bondi radius from an inner $\varrho \propto R^{-3/2}$ region to an outer one with a fast convergence to the flat asymptote, after
a few tens of Bondi radii, further changes in the density profile are below the percent level. The accretion of hot gas by a SMBH is a very distinct
situation, where no infinite gas cloud of constant density is in evidence. The presence of the central black hole is in now way a minor perturbation on
the structure of a constant density distribution and there is no large radius constant density distribution to be seen. Indeed, it is clear that
in the absence of the central black hole, the gas distribution would be significantly different. Thus, the accretion density profile is determined
by the central potential out to large radii, and this density profile continues to fall outside of the Bondi radius by many orders of magnitude out
to a large radial dynamical range and down to densities smaller than the values at the Bondi radius by many orders of magnitude. The many observational
studies quoted above make it clear that the solution of eqs.(13)-(15) represents a much closer description to reality than the classical Bondi model, at
least for the cases of hot accretion onto SMBHs.

We complete the description of the spherically symmetric solution of eqs.(13)-(15) in {  appendix B} at the end of the paper where
we show that with the exception of the region about $\mathcal{M}=1$, small deviations about $\gamma=5/3$ result only in small changes in the
power law indices describing the steady state spherical solution of eqs.(13) and (15).

\section{Deviations from spherical symmetry}

We now use the solution derived in the previous section as a ground state for a first order perturbation analysis of deviations from
sphericity in the density distribution, angular momentum and velocity fields. {  Given the simplicity of the accretion solution of equations
(13)-(15), this development will be much simpler than the corresponding exploration of non-spherical deviations for the Bondi model, e.g.
Foglizzo \& Ruffert (1996), Foglizzo (2001). }

We shall go back to the system of equations (7-10), assuming
the generic solution:

\begin{equation}
  \mathcal{V}=\mathcal{V}_{0}R^{-1/2}+\epsilon\mathcal{V}_{R}(R)\mathcal{V}_{\theta}(\theta),
\end{equation}

\begin{equation}
 \mathcal{U}=\epsilon\mathcal{U}_{R}(R)\mathcal{U}_{\theta}(\theta),
\end{equation}

\begin{equation}
 \mathcal{W}=\epsilon \mathcal{W}_{R}(R)\mathcal{W}_{\theta}(\theta),
\end{equation}

\begin{equation}
  \varrho=\varrho_{0} R^{-3/2} \left(1+ \epsilon \varrho_{R}(R)\varrho_{\theta}(\theta) \right),
\end{equation}

\noindent where $\epsilon <<1$, a separation of variables ansatz for the non-spherical perturbation e.g. Bender \& Orszag (1978).
{  The assumption of a barotropic adiabatic $\gamma=5/3$ equation of state would be invalidated
by the presence of shocks, viscosity, dissipation or crossing streamlines, neither of
which are present in our model or appear in any of the solutions found. The lack of
spherical symmetry or the number of spatial dimensions defining the flow solution imply
no thermodynamic constraints in and of themselves, and hence in no way affect the
validity of the $\gamma=5/3$ assumption. On this, see e.g. Chapter 4 in Clarke and Carswell (2007) or
Gruzinov (2022) for a recent example of the use of $\gamma=5/3$ in the study of astrophysical accretion outside of spherical symmetry,
or Mancino (2022) who consider $\gamma=5/3$ whilst studing spherical accretion about a black hole.}

When introducing the above ansatz into equations (7-10), the $\varrho^{-1/3}$ non-linearities in equations (8) and (9)
will be treated through a series expansion keeping only terms to first order in $\epsilon$. The order $\epsilon^{0}$
terms will collectively sum to zero in the {  four} equations, as they satisfy the unperturbed spherical system, while
all terms to order $\epsilon^{2}$ and higher appearing in eqs. (7-10) will be dismissed. {  The first point to notice is that the
azimuthal velocity, $\mathcal{W}$, does not appear in the mass conservation equation, eq.(7), and enters only quadratically in eqs. (8) and (9).
Hence, to first order in $\epsilon$, equations (7-9) form a system of three equations in the three perturbed non-spherical components
of the radial and polar velocity and density fields. Then, eq.(10) constraints $\mathcal{W}$. We shall now solve the system of eqs.(7-9)
for the perturbations in $\mathcal{V}_{R}(R)\mathcal{V}_{\theta}(\theta)$, $\mathcal{U}_{R}(R)\mathcal{U}_{\theta}(\theta) $ and
$ \varrho_{R}(R)\varrho_{\theta}(\theta) $, leaving the simpler case of the constraint on $\mathcal{W}$ through eq.(10) for later.}

Equation (9) yields:

\begin{equation}
  \epsilon \mathcal{V}_{0} \mathcal{U}_{\theta} R^{1/2} \frac{d \mathcal{U}_{R}}{d R}+\epsilon \mathcal{V}_{0} \mathcal{U}_{\theta}
  \mathcal{U}_{R} R^{-1/2}=-\epsilon \varrho_{0}^{2/3} R^{-1} \varrho_{R} \frac{d \varrho_{\theta}}{d \theta},
\end{equation}

\noindent which can be simplified to:

\begin{equation}
  \frac{R^{1/2}}{\varrho_{R}} \frac{d(R \mathcal{U}_{R})}{dR}=-\frac{\varrho_{0}^{2/3}}{\mathcal{V}_{0}\mathcal{U}_{\theta}}
  \frac{d \varrho_{\theta}}{d \theta}.  
\end{equation}

The above equation yields two conditions:

\begin{equation}
C_{3} \varrho_{R}= R^{1/2} \frac{d(R\mathcal{U}_{R})}{dR},
\end{equation}

\noindent and

\begin{equation}
  \varrho_{0}^{2/3} \frac{d\varrho_{\theta}}{d\theta}=-C_{3} \mathcal{V}_{0} \mathcal{U}_{\theta},
\end{equation}

\noindent where $C_{3}$ is a separation constant.

Equation (8) yields:

\begin{equation}
\begin{split}
  \epsilon \mathcal{V}_{0} \mathcal{V}_{\theta} \left( R^{1/2} \frac{d \mathcal{V}_{R}}{dR}-
  \frac{\mathcal{V}_{R}}{2 R^{1/2}} \right) =\\
  -\epsilon \varrho_{0}^{2/3} \varrho_{\theta} \left[
  \frac{\varrho_{R}}{2R} + R^{3/2} \frac{d(\varrho_{R}/R^{3/2})}{dR} \right], 
\end{split}
\end{equation}

\noindent which can be simplified to:

\begin{equation}
  \mathcal{V}_{0}\mathcal{V}_{\theta} \frac{d(\mathcal{V}_{R}/R^{1/2})}{dR}=
  -\varrho_{0}^{2/3} \varrho_{\theta}  \frac{d(\varrho_{R}/R)}{dR}.
\end{equation}


The above equation yields two conditions:

\begin{equation}
   \frac{d(\mathcal{V}_{R}/R^{1/2})}{dR}=C_{2} \frac{d(\varrho_{R}/R)}{dR},
\end{equation}


\noindent and,

\begin{equation}
   \varrho_{\theta}=-\left( \frac {C_{2}\mathcal{V}_{0}}{\varrho_{0}^{2/3}} \right) \mathcal{V}_{\theta},
\end{equation}

\noindent where $C_{2}$ is a separation constant. 

Finally, equation (7) yields:

\begin{equation}
  \epsilon \varrho_{0} \mathcal{V}_{\theta} \frac{d(\mathcal{V}_{R}R^{1/2})}{dR}+
  \epsilon \varrho_{0} \mathcal{V}_{0} \varrho_{\theta} \frac{d \varrho_{R}}{dR}=
  -\frac{\epsilon \varrho_{0}\mathcal{U}_{R}}{R^{1/2} \sin \theta} \frac{d(\mathcal{U}_{\theta} \sin \theta)}{d \theta}.
\end{equation}

Although the above equation is not immediately susceptible to separation of variables, use of eq.(33) allows to replace
$\mathcal{V}_{\theta}$ for $\varrho_{\theta}$ and hence yields the following two conditions:

\begin{equation}
  \mathcal{V}_{0} \frac{d \varrho_{R}}{dR} -\frac{\varrho_{0}^{2/3}}{C_{2}\mathcal{V}_{0}}\frac{d(\mathcal{V}_{R}R^{1/2})}{dR}=
  \frac{C_{1}\mathcal{U}_{R}}{R^{1/2}},
\end{equation}

\noindent and,

\begin{equation}
\frac{d \mathcal{U}_{\theta}}{d \theta}+\cot \theta\mathcal{U}_{\theta}=-C_{1}\varrho_{\theta},
\end{equation}

\noindent where $C_{1}$ is a separation constant. We are hence left with a purely radial system in exact radial derivatives of eqs.(28),
(32) and (35), and a purely angular system in exact angular derivatives of eqs.(29), (33) and (36), which we shall deal with first.
Equation (29) shall be used to replace $\mathcal{U}_{\theta}$ in eq.(36) for derivatives of $\varrho_{\theta}$, yielding an equation in
this last variable alone:

\begin{equation}
  \frac{d^{2} \varrho_{\theta}}{d \theta^{2}} +\cot \theta \frac{d\varrho_{\theta}}{d\theta}
  -\left( \frac{C_{1}C_{3} \mathcal{V}_{0}}{\varrho_{0}^{2/3}} \right)\varrho_{\theta}=0.
\end{equation}

This last equation has as solutions Legendre polynomials of $\cos \theta$ of the first and second kinds, and hence opens a very
rich phenomenology capable of modeling a wide variety of density configurations, and given the constraints linking $\mathcal{U}_{\theta}$
and $\mathcal{V}_{\theta}$ to $\varrho_{\theta}$, an extensive range of flow patterns. Considering for example odd polynomials, or the addition
of odd and even polynomials, will yield asymmetric velocity fields. This last might even suggest the existence of an intrinsic component
to observed morphological and energetic jet asymmetries such as the ones often observed in FR II systems e.g., Hardcastle et al. (1999) or
Liu et al. (2020). As the Legendre polynomials form an orthogonal basis, any given axisymmetric flow configuration can be reproduced
with the velocity solutions presented here, with the model then yielding a corresponding self-consistent density configuration. Similarly,
any desired density configuration within the limits of the quasi-spherical approximation, can be obtained as a sum of solutions to eq.(37),
with the model then yielding a corresponding self-consistent flow configuration.

In this first exploration of the problem, and guided by the simple bipolar infall/outflow geometries observed in the numerical experiments of
Aguayo-Ortiz et al. (2019), Tejeda et al. (2020) and Waters et al. (2020), we choose to consider here only solutions of the type:

\begin{equation}
-C_{1}C_{3} \mathcal{V}_{0}/\varrho_{0}^{2/3}=6,
\end{equation}
  
\noindent which will yield a simple geometry with $\mathcal{U}_{\theta}$ satisfying the boundary conditions
$\mathcal{U}_{\theta}(0)=\mathcal{U}_{\theta}(\pi/2)=0$, 

\begin{equation}
  \varrho_{\theta}=\varrho_{1}\left(3\cos^{2}\theta-1 \right),
\end{equation}

\begin{equation}
  \mathcal{V}_{\theta}=-\varrho_{1} \left( \frac{\varrho_{0}^{2/3}}{C_{2}\mathcal{V}_{0}}  \right) \left( 3\cos^{2}\theta-1  \right),
\end{equation}

\begin{equation}
   \mathcal{U}_{\theta}= \varrho_{1}  \left( \frac{6\varrho_{0}^{2/3}}{C_{3}\mathcal{V}_{0}}  \right)\cos \theta \sin \theta,
\end{equation}

\noindent which solve the angular system in terms of an amplitude constant $\varrho_{1}$. {  We stress that the particular choice of Legendre
polynomial adopted at this point is arbitrary and is motivated merely by mathematical convenience and the qualitative correspondence of the
resulting velocity flow patters to those of the previous numerical experiments mentioned above.}

We now turn to the radial system of eqs.(28), (32) and (35), which can be solved taking radial power laws for the unknown functions:

\begin{equation}
\mathcal{V}_{R}=R^{\alpha}, \mathcal{U}_{R}=R^{\beta} , \varrho_{R}=R^{\delta},
\end{equation}

\noindent where $\alpha$, $\beta$ and $\delta$ are real. Equation (28) becomes:

\begin{equation}
  (\beta +1)R^{\beta}=C_{3}R^{\delta-1/2},
\end{equation}

\noindent which yields conditions:

\begin{equation}
\beta=\delta-1/2,  
\end{equation}

\noindent and,

\begin{equation}
\beta+1=C_{3}.  
\end{equation}

Equation (32) becomes:

\begin{equation}
  (\alpha -1/2)R^{\alpha-3/2}=C_{2}(\delta-1)R^{\delta-2},
\end{equation}

\noindent which yields conditions:

\begin{equation}
\alpha=\delta-1/2  \Rightarrow \alpha=\beta, 
\end{equation}

\noindent and,

\begin{equation}
C_{2}=1.  
\end{equation}

Lastly, equation (35) becomes:

\begin{equation}
  \delta \mathcal{V}_{0}R^{\delta-1}-\frac{\varrho_{0}^{2/3}}{\mathcal{V}_{0}}(\alpha+1/2)R^{\alpha-1/2}=C_{1} R^{\beta-1/2}
\end{equation}

\begin{figure}
\hskip -10pt \includegraphics[width=9.25cm,height=8.0cm]{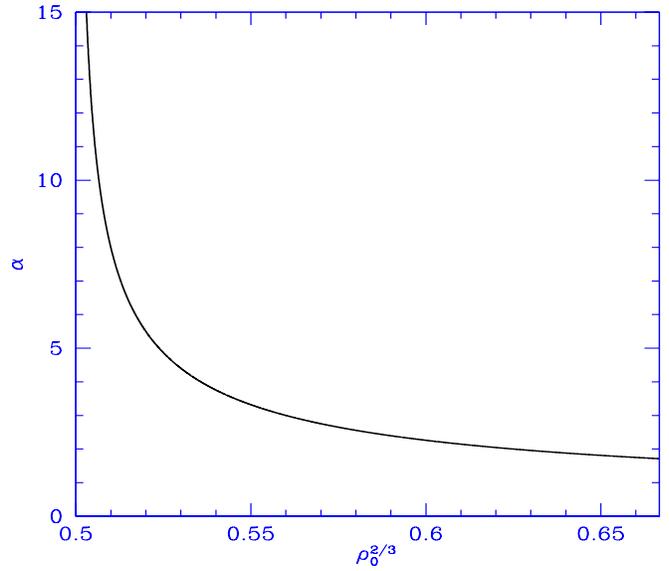}
\caption{The curve gives the value of the radial power law indices $\alpha=\beta$ for the non-spherical perturbative
  solution and the '+' sign in eq.(53), as a function of the dimensionless density parameter of the spherically symmetric one,
  $\varrho_{0}^{2/3}$. This diverges as one approaches the maximum accretion rate value at $\varrho_{0}^{2/3}=1/2$.}
\end{figure}

\noindent which yields constraints on the power indices consistent with those obtained previously, and:

\begin{equation}
  \delta \mathcal{V}_{0}-\frac{\varrho_{0}^{2/3}}{\mathcal{V}_{0}}(\alpha+1/2)=C_{1}.
\end{equation}

\noindent We can now solve for the power indices in terms of the unique free parameter of the unperturbed solution
$\varrho_{0}$, by replacing $\alpha$ for $\delta-1/2$ and $C_{1}$ for its value through condition (38) into the above
equation to obtain an equation in $\delta$ alone:

\begin{equation}
\delta(\delta+1/2)\left(1-\frac{\varrho_{0}^{2/3}}{\mathcal{V}_{0}^{2}} \right)=-6\frac{\varrho_{0}^{2/3}}{\mathcal{V}_{0}^{2}},  
\end{equation}

\noindent which can be written as:

\begin{equation}
\delta(\delta+1/2)=A,
\end{equation}

\noindent where $A=3\varrho_{0}^{2/3}/(2\varrho_{0}^{2/3}-1)$, having solutions:
\begin{equation}
\delta=-\frac{1}{4}\pm \left(\frac{1}{16}+A \right)^{1/2},  
\end{equation}

\noindent which completes the full solution in terms of the unique parameter of the unperturbed model $\varrho_{0}$ as:

\begin{equation}
\varrho=\varrho_{0}R^{-3/2}\left(1+\epsilon \varrho_{1} R^{\delta}\left[3 \cos^{2} \theta -1  \right]   \right),
\end{equation}

\begin{equation}
  \mathcal{V}=\mathcal{V}_{0}R^{-1/2}- \epsilon \varrho_{1}  R^{\delta-1/2}\left(\frac{\varrho_{0}^{2/3}}{\mathcal{V}_{0}}
  \right)\left[3\cos^{2} \theta-1  \right],
\end{equation}

\begin{equation}
  \mathcal{U}= \epsilon \varrho_{1}  R^{\delta-1/2} \left( \frac{6\varrho_{0}^{2/3}}{[\delta+1/2]\mathcal{V}_{0}}
  \right)\cos \theta \sin \theta.
\end{equation}

\noindent Equation (53) has real roots only for the sub-sonic regime of $\varrho_{0}^{2/3}>1/2$, where one root is larger than zero,
and the small interval $0< \varrho_{0}^{2/3} <0.02$, where both roots are smaller than zero. The above ranges of $\varrho_{0}$ hence
define validity limits for the perturbative analysis presented for the radial system. Figure (5) gives a plot of the $\alpha$ radial power-law index as a
function of $\varrho_{0}^{2/3}$ for the '+' sign in eq.(53), for the subsonic region of parameter space. We see a mild variation, with
the exception of a highly localised divergence towards $\mathcal{M}=1$. Notice that for $\alpha>-1/2$, $\delta>0$ and hence the spherically
symmetric components of the velocity and density fields dominate over the non-spherical ones as $R \to 0$, leaving the accretion rate
unchanged with respect to the one calculated in the previous section, which assumed spherical symmetry.

{  Finally, we turn to eq.(10) as a constraint on $\mathcal{W}$, which to first order in $\epsilon$ simply yields:

\begin{equation}
\frac{\partial \mathcal{W}_{R}}{\partial R}=-\frac{\mathcal{W}_{R}}{R},
\end{equation}

\noindent and hence,

\begin{equation}
\mathcal{W}_{R}=\mathcal{W}_{0}/R,
\end{equation}

\noindent where $\mathcal{W}_{0}$ is a scaling constant. Thus we see that the perturbation in the azimuthal velocity field is such that
the specific angular momentum, $l=\mathcal{W}_{R}\mathcal{W}_{\theta}R \sin \theta$, does not depend on the radial coordinate. Indeed, the
assumption of a conserved specific angular momentum, $l$, for the problem now fixes $\mathcal{W}_{\theta}=\sin^{-1} \theta$.
{  The specific form of $\mathcal{W}_{\theta}$ found above violates the perturbative introduction of angular momentum presented here
  interior to a critical radius and angle combination, as towards the $\sin \theta=0$ poles $\mathcal{W}_{\theta}$ diverges. This shows
  that introducing any amount of angular momentum, which implies a centrifugal barrier, becomes inconsistent with the linear perturbation
  analysis of this section interior to some critical inner radius. Still, given the quadratic appearance of the small parameter $\epsilon$
  in all azimuthal velocity terms, this will naturally be a second-order effect relevant only when considering a non-zero
  level of angular momentum.}

The assumption of $\epsilon<<1$ which characterises the perturbative analysis presented, defines the validity range of the results, firstly
though the quasi-spherical series used in developing the non-linear density terms in eqs. (8) and (9). Hence, a first limit appears
whenever the non-spherical term in the density field in eq (54) becomes comparable to the spherical one. Then, the same applies to the
velocity field, in particular, for the radial velocity of eq.(55). Whenever the non-spherical term becomes comparable to the spherical
one, the validity limit of the mathematical development is reached and there is no certainty that the solution remains a good description
of the physical situation being modelled.

However, as discussed in the following section, it is interesting that within the validity range of the density series expansion
we see that the velocity field remains a qualitative match to the numerical velocity fields of Aguayo-Ortiz et al. (2019) and Waters
et al. (2020) even after the non-spherical term in the velocity field dominates over the spherical one. {  Indeed, a first series
of numerical experiments presented in section 5 confirm that the convergence steady state solution to initial conditions given by
the equatorial infall and polar outflow modes of eqs.(53)-(56), preserves the qualitative structure of the flow, although quantitative
changes appear in the details as one goes beyond the validity regime of the linear approximation.}

For the positive root of
$\delta$, towards the hydrostatic equilibrium limit, small $\mathcal{V}_{0}$ values can yield an outer region where the second term
in eq.(55) dominates over the radial infall of the first term in this equation (as $\mathcal{V}_{0}<0$), and lead to the development
of a bipolar outflow highly reminiscent of the results of the numerical studies mentioned above.

Regarding the physical conditions which might be responsible for the boundary conditions required for any of the many options allowed by
the model, any small amplitude, large angular scale departure from spherical symmetry in density and/or velocity could be described by a sum
of Legendre functions solving eq.(37). We chose not to focus on any specific astrophysical situation as a way to stress the generality of
the model presented, which we believe can be of relevance over a variety of scenarios. {  However, in the context of the hot x-ray gas
accreting onto SMBHs discussed in section 2 and appendix A, the presence of a small degree of angular momentum would imply the appearance of
large angular scale, small amplitude non-spherical modes in the accreting material. This would become particularly relevant towards the
innermost regions, probably playing a part in the initial jet launching dynamics, as suggested by the equatorial infall/polar outflow solutions
discussed in the following section.}

Since the mass accretion rate is not modified, the material in the polar outflows is hence supplied by an infall rate above the one of the
spherically symmetric solution. As seen also in the choked accretion results of Aguayo-Ortiz et al. (2019) and Tejeda et al. (2020),
it is the non-spherically symmetric accretion of material above the spherically symmetric accretion rate, which results in the
bipolar outflows obtained. Thus, our present results qualitatively maintain the choked accretion phenomenology described
in the two references above.

\section{Particular examples}

\begin{figure*}
  \hskip -20pt \includegraphics[height=8cm,width=9.8cm]{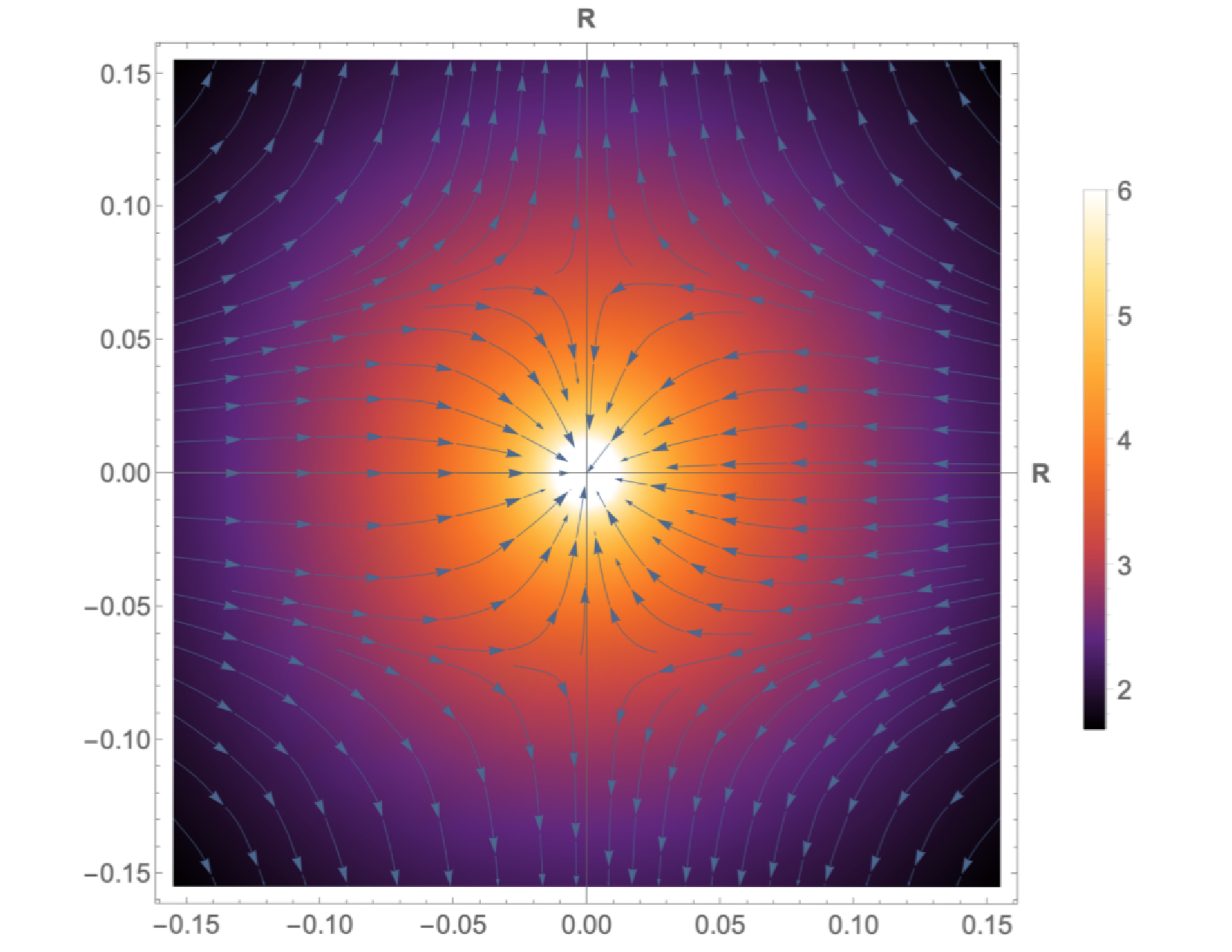}
  \includegraphics[height=8cm,  width=\columnwidth ]{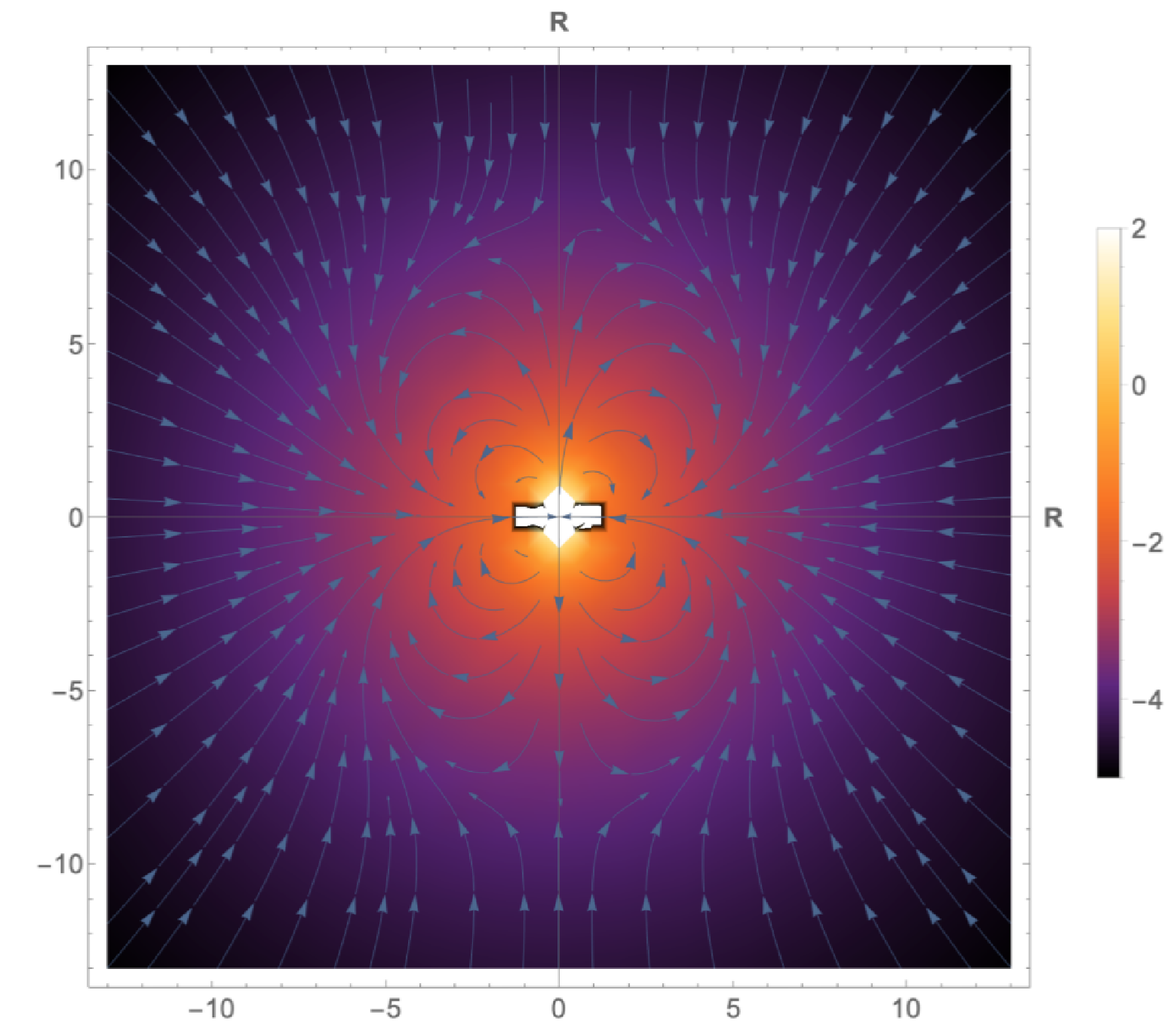} 
  \caption{{  Left}:Velocity streamlines for the example described in the text taking the positive root in eq.(53), overlayed on the
    logarithm of the corresponding density field, which can be seen to deviate from sphericity only minimally. A stagnation point is
    evident along the polar axis at $R_{J}=0.073$. We see an inner spherically symmetric infall region evolving outwards into an outer
    region with a polar outflow and equatorial infall. {  Right}: Velocity streamlines for the example described in the text taking the
    negative root in eq.(53), overlayed on the logarithm of the corresponding density field, which can be seen to deviate from sphericity
    only minimally. The stagnation point is evident along the polar axis at $R_{J}=8.4$. We see an outer spherically symmetric infall region
    evolving inwards into an inner region with a polar outflow and equatorial infall.}
\end{figure*}

We now present a particular example of a model close to the hydrostatic equilibrium limit of $\varrho_{0}^{2/3}=2/3$, taking
$\varrho_{0}^{2/3}=0.66533$ which results in $\mathcal{V}_{0}^{2}=0.004$ and $\delta=2.22$, $\alpha=1.72$, for the '+' sign in
equation (53).

A plot of the velocity flow of this model is shown in the left panel of fig.(6), superimposed on the logarithm of the density field
shown as a colour scale plot. The equatorial infall and polar outflow structure is evident at large radii, together with the highly
spherical configuration of the density field. This last however, is not exactly spherical, it is the slight prolateness present that
sources the non-radial terms in radial and angular velocities.

The velocity flow of the left panel in fig.(6) is qualitatively equivalent to those which arise from the numerical
simulations of Aguayo-Ortiz et al. (2019),  Waters et al. (2020) and Tejeda et al. (2020), showing an inner region tending to
spherical accretion in both density and velocity fields, and a more external equatorial infall and polar outflow one. {  A first
quantitative comparison of our analytic solution to numerical simulations appears in section 5.} In the case
shown here, a small value of $\mathcal{V}_{0}$ results in strong polar outflows with radial velocities which become larger than the
local escape velocities well within the validity regime of the density quasi-spherical approximation.

It is clear that the velocity outflow described above is far from the narrow collimated jets of astrophysical objects, where
other physics such as magnetic fields and rotation clearly play a role. However, the model described could present a complementary
ingredient in helping to initially turn an infall into an outflow, or to describe the much wider-angle components inferred sometimes
to accompany some jet phenomena  (e.g. Sato et al. 2021 or Duque et al. 2022).

%
%

As can be seen from eq.(55), the radial velocity of the ejected material is
a strong function of the angle, decreasing gradually at first for small angles, but then at an increasing rate, reminiscent of the
structured jets inferred in some GRB studies (e.g. the Gaussian angular velocity profiles of jets treated in Lamb et al. 2021 or the
structured jets of Kathirgamaraju et al. 2018), or of the transition towards an isotropic phase envisioned for GRBs, e.g. Waxman (2004).

Although the results described above occur within the validity regime of the density approximation, on reaching the point where the
perturbative velocity component becomes comparable to the spherically-symmetric one, the formal validity regime of the scheme developed
is reached. Hence, we cannot be certain that the extrapolation of the model is valid out to the regime where the radial infall velocity
becomes an outflow.



Given the qualitative agreement of the solution presented to the various numerical experiments described above, we can present a first quantitative
comparison in terms of the most general feature of the solution shown above. It is clear from eq.(55) that as $R \to \infty$, the
radial velocity will have a zero where $3 \cos^2 \theta =1$ , $\theta=54.74^{\circ}$. This can be compared to the published numerical results of
Aguayo-Ortiz et al. (2019) (fig 4, bottom right) and  Waters et al. (2020) (e.g. fig 1 leftmost panel) where the angle at which the radial
velocity equals zero at the outer edge of their simulation domains (and hence a lower value for the final asymptotic value of this parameter)
is of 51.3 and 52.2 degrees, respectively. We hence see that the numerical results listed above are {  not inconsistent with the value of the angle
for which the radial velocity of eq.(55) goes to zero for large radii}. Indeed, in the first numerical experiments presented in section 5
we see that numerical convergence beyond the validity range of the linear approximations of the previous section yield solutions which maintain
the qualitative structure of the model presented in sections 3 and 4.

Lastly, we look at the minus sign in eq.(53), which now fixes a value of $\delta=-2.72$, for the same choice of $\varrho_{0}^{2/3}=0.66533$
as taken above. This solution is qualitatively different to the previous for a number of reasons. First, we see from the denominator
in eq.(56) that the sign of the angular velocity will change, meaning that instead of focusing the flow towards the poles, this will happen
towards the equator. Also, as the radial powers of the perturbative terms are now smaller than the ones of the spherically symmetric
terms in eqs.(54) and (55), the unperturbed solution will dominate for large radii, and the non-spherical terms become relevant at
small, rather than large values.

In the right panel of fig.(6) we present the velocity flow of this last example, together with the logarithm of the density field; this last,
as in the case of the left panel in this figure, deviates only marginally from spherical symmetry. The dominance of the unperturbed spherical
solution of eqs.(13) and (15) is evident towards the outer regions of the figure, as is the change in the inflow solution for small radii.
This transition occurs well within the density validity limit of the solution for small radii, denoted by the exclusion rectangle
at the centre. For small radii we see the appearance of a radial outflow of limited extent, and of the funnelling of the accreted material
onto the equatorial plane, at a strongly centrally divergent velocity.

The above results are interesting in light of recent indications of super-Eddington accretion in a variety of astrophysical scenarios.
The discovery of over 200 quasars at very early redshifts $z>6$ (e.g. Mortlock et al. 2011, Ba\~{n}ados et al. 2018) imply black hole growth
rates well above the Eddington limit, e.g. Volontieri et al. (2015). At stellar scales, X-ray binaries (e.g. Okuda 2002) or ultra Luminous
X-ray sources inferred to contain black holes (e.g. Winter et al. 2006) have also been interpreted as evidence for super-Eddington accretion.
At a fundamental level, exceeding the Eddington limit is a natural consequence of the breaking of spherical symmetry in the accretion flow,
e.g. Paczynski \& Abramowicz (1982), Massonneau et al. (2022). As such, the accretion model presented in the right panel of fig.(6), does
precisely that; a large-scale spherically-symmetric accretion flow is redirected into an equatorial infall covering a very small solid angle,
presenting a minimal interaction cross-section to any photon flux coming from the central object.

As happens also with the case shown in the left panel of fig.(6), the solution described lies within the validity regime of the density
approximation beyond the central exclusion rectangle. However, the formal validity regime of the velocity approximation ends before the point
where the radial velocity changes sign. Thus, the equatorial focusing described is only a mild effect within the full validity regime of the
solution.  Over said range this equatorial focusing is seen as a monotonically growing effect, highly suggestive of this becoming dominant at
smaller radii, which unfortunately cannot be traced reliably with the perturbative scheme developed.

The examples shown above might seem to be artificially close to the hydrostatic equilibrium $\varrho_{0}^{2/3}=2/3$ limit. However, during
sufficiently early phases of massive stellar collapse, for example, the system can be chosen arbitrarily close to hydrostatic equilibrium.
As can be seen from eq.(55), as $\mathcal{V}_{0} \to 0$ towards the hydrostatic equilibrium limit, the polar positive outflow at $\theta=0$
diverges. Note also that the model presented is equally applicable to accretion about a black hole, provided the region of interest
remains at radii much larger than the Schwarzschild radius of the problem, and hence within the Newtonian validity regime.

\begin{figure*}
  \hskip -20pt \includegraphics[height=8cm,width=8.5cm]{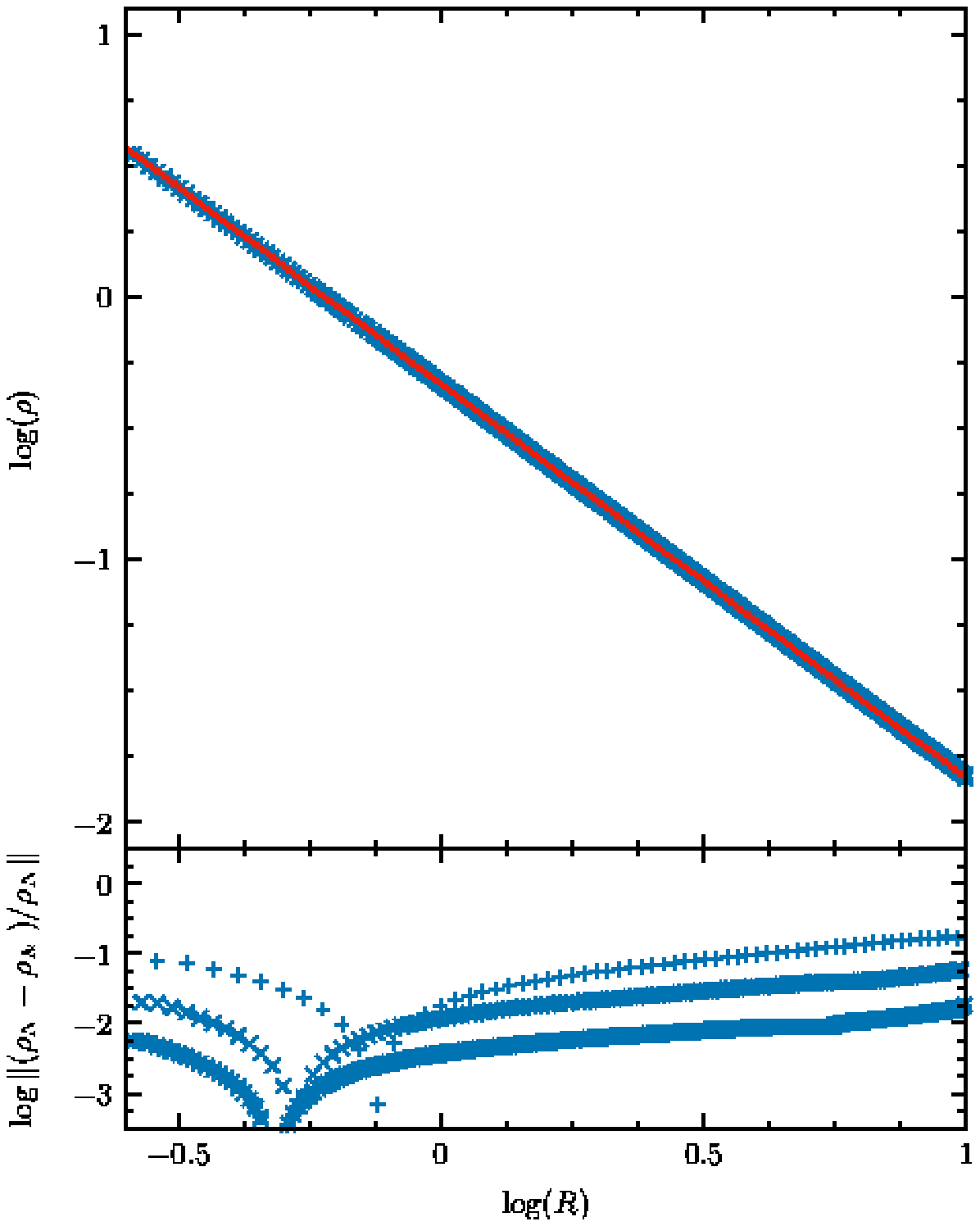}
  \includegraphics[height=8cm,  width=8.5cm ]{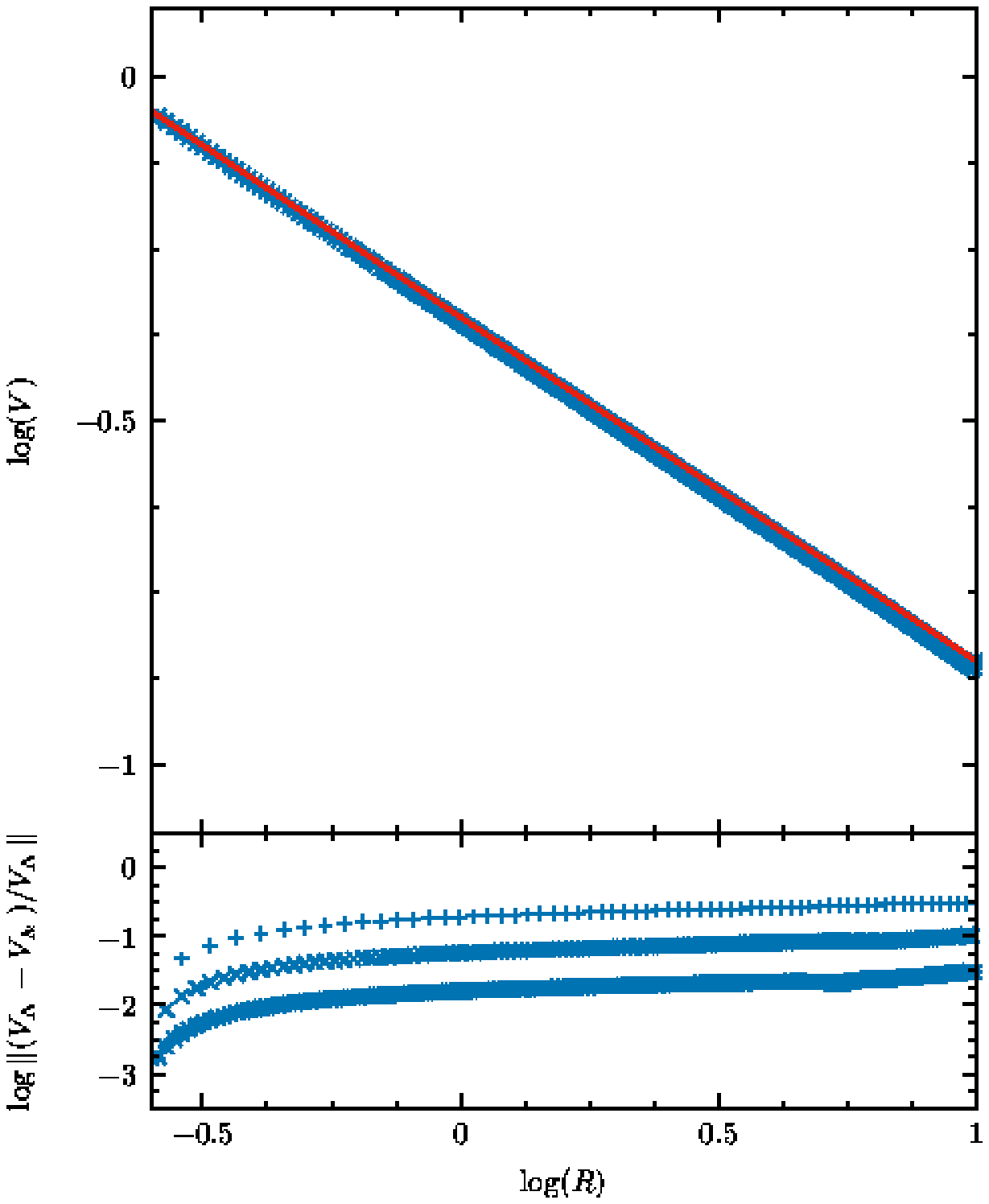} 
  \caption{{Left}: Top panel shows numerical convergence density field solutions for initial conditions as given by the exact solution of
    eqs. (13) and (14), for a sub0sonic case of $\varrho^{2/3}=0.6$, blue crosses, together with the exact solution, red line.
    The bottom panel gives the logarithm of the relative error between the exact solution and numerical converge 1D solutions
    for radial resolutions of 64, 128 and 256, showing the numerical convergence of the simulations. 
    {Right}: We here show the corresponding velocity field results, for the same simulations appearing in the left panel.}
\end{figure*}

\section{First numerical explorations}

{ 
In this final section we present a numerical setup to be used in the exploration of the results of the previous sections,
showing at this point the results of only the first of many computational experiments relevant to the new spherical accretion
solution of section 2 and its non-spherical perturbations as described in section 3. 

We perform numerical simulations using the hydrodynamical code {\sc{aztekas}} (Aguayo-Ortiz \& Mendoza 2021), which solves the
inviscid Euler equations in a conservative form, coupled with an ideal gas equation of state providing a polytropic relation for
the pressure. A detailed description of the code and extensive validation against standard analytical hydrodynamical tests can be
found in Aguayo-Ortiz et al. (2018) and Tejeda \& Aguayo-Ortiz (2019). All the simulations presented here use a High Resolution
Shock Capturing scheme along with a second-order piecewise linear reconstructor and an HLL approximate Riemann solver. The time
evolution is computed using a second-order total variation diminishing Runge-Kutta time integrator, with a Courant factor of 0.5.

We first reproduce the new spherically-symmetric accretion solution of eqs.(13) and (14), by choosing a $\varrho^{2/3}=0.6$, corresponding
to a case in the sub-sonic region, and using $\gamma = 5/3$. A Keplerian potential due to a central point mas $M$ is assumed and the
equations are all written in dimensionless variables such that $G=M=1$, following also the dimensionless variables used in section 2.
1D spherically-symmetric simulations were performed using a domain consisting of 64, 128 or 256 numerical cells  distributed over an exponential
radial grid with $R \in [0.25,10]$, as in Aguayo-Ortiz et al. (2019). Initial conditions were chosen by taking the analytic solution of
eqs. (13) and (14). Both the inner and outer boundary conditions were fixed at the values given by the analytic solution at all times,
while the solution for the rest of the domain was completely free to evolve and settle to whichever radial profiles it naturally converged.
Convergence is defined as the point where the accretion rate, averaged over the entire volume, has a relative change between subsequent
timesteps of less than one part in $10^{6}$.

The results of one such simulation are given in Figure 7, where the top panels show the density and velocity radial profiles at convergence,
left and right panel respectively, for the three resolutions mentioned above, blue crosses, together with the initial conditions, the analytic solution
of eqs. (13) and (14), solid red line. It is clear that the numerical convergence solution very closely matches the new analytic solution. Indeed,
the bottom panels of this figure give the logarithm of the deviation between the convergence numerical solution and the exact analytic solution,
for runs using radial resolutions of 64, 128 and 256, top to bottom of these panels. Numerical convergence is evident, as the fractional error
rapidly falls as the resolution increases. Cases with $\varrho^{2/3}<0.5$, corresponding to the super-sonic region of the spherical solution,
were also tested and give similar results, a quick convergence to the analytic model, with even greater robustness that for the sub-sonic case.
As in that case perturbations do not propagate outwards, the inner boundary conditions can be left free. This tests {  show the stability of
the new solution of eqs. (13) and (14),} as numerical instabilities are seen not to propagate or grow as the numerical solution converges
quite rapidly to the exact solution of these eqations. As both the density and infall velocity radial profiles converge to the analytic
exact solution of eqs. (13) and (14), given the equation of state taken, the flat profile of $\mathcal{M}$ is also verified in the simulations.

\begin{figure*}
  \hskip -20pt \includegraphics[height=8cm,width=8.5cm]{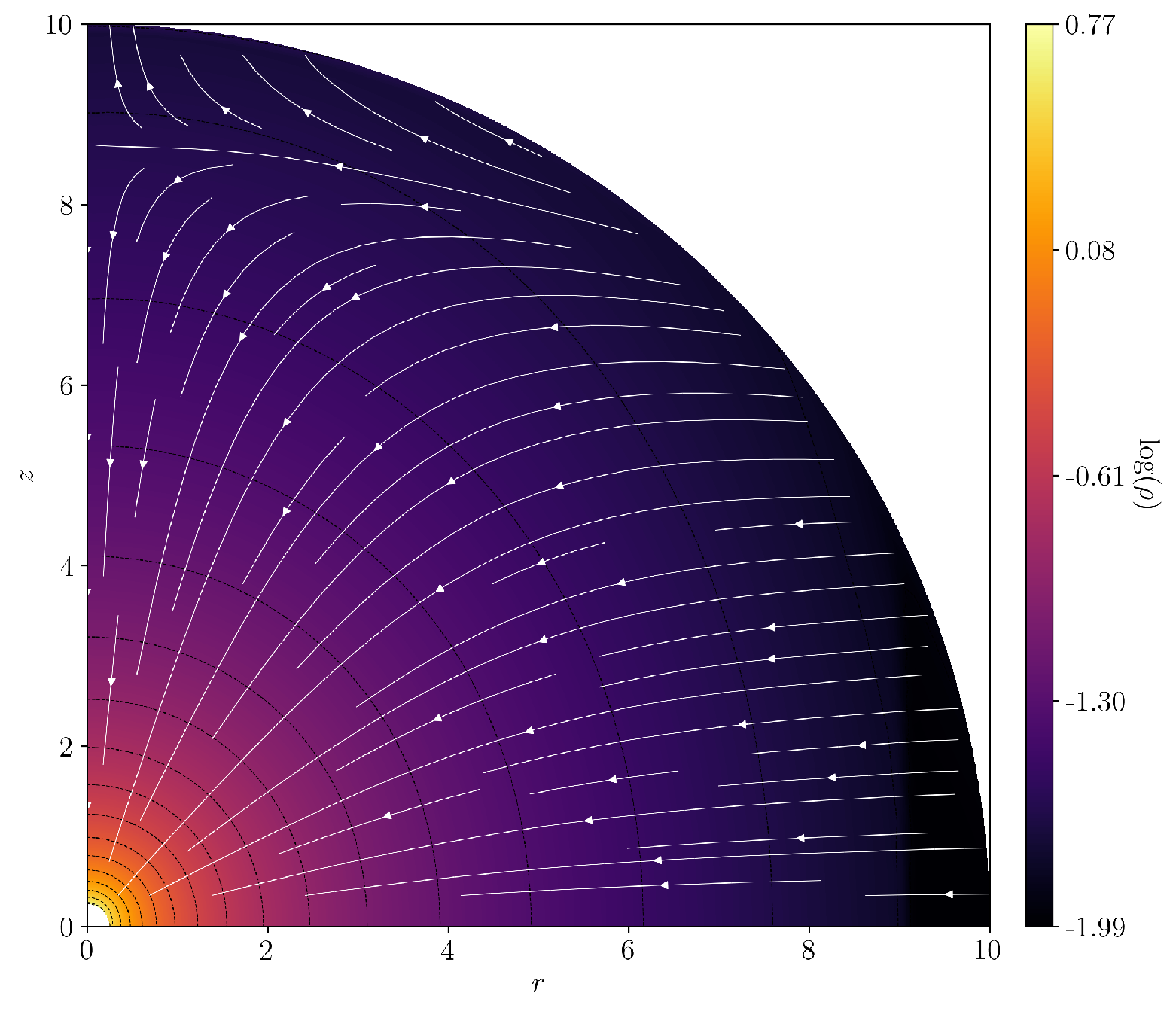}
  \includegraphics[height=8cm,  width=8.5cm]{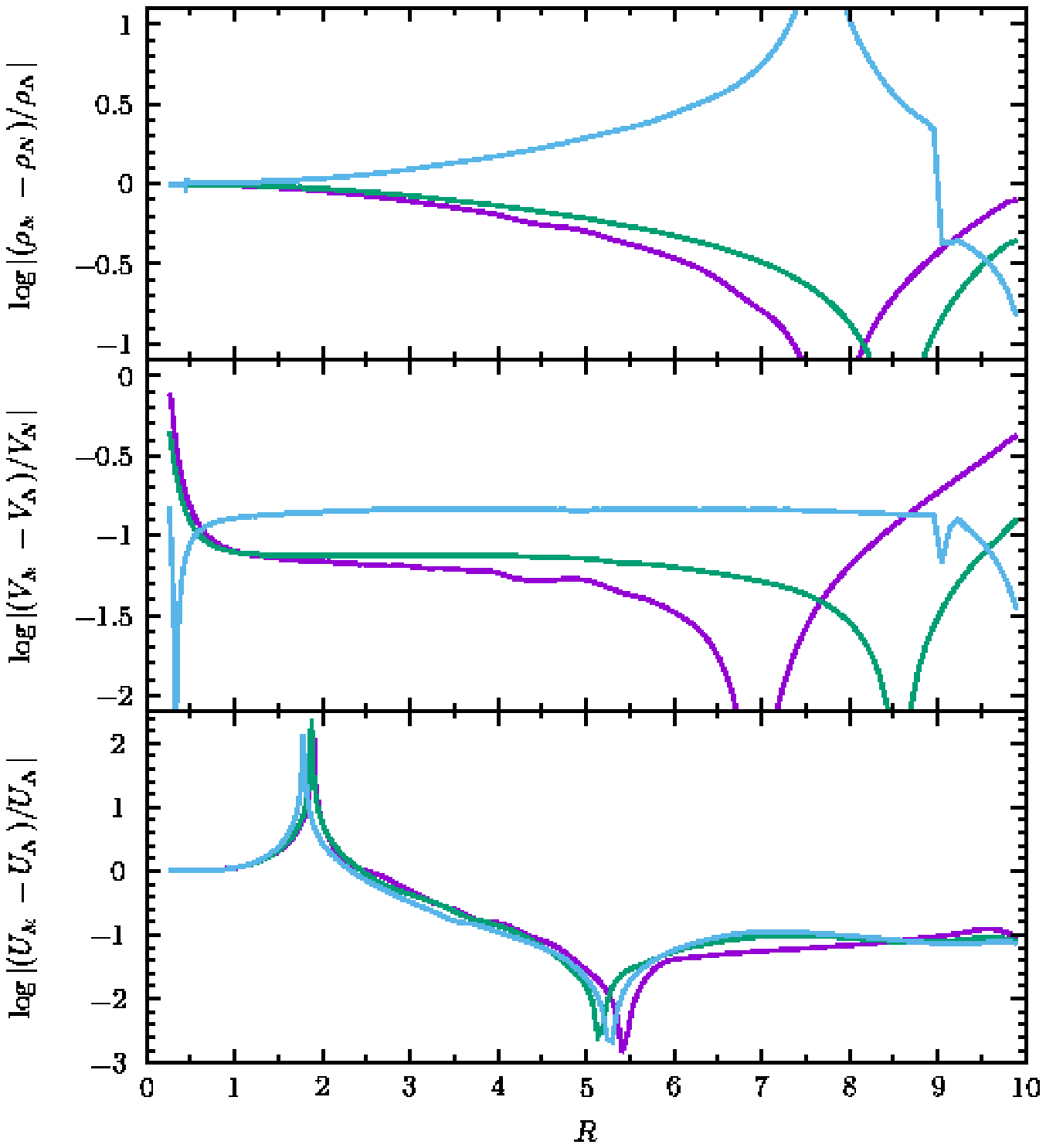} 
  \caption{{Left}: Final convergence 2D numerical results, velocity fields and density field for initial conditions as given by
    the linear perturbed solution of eqs. (53)-(56), for $\varrho^{2/3}=0.56$, $\epsilon \varrho_{1}=0.0005$ and taking the plus sign in eq.(53).
    The qualitative character of the initial equatorial infall / radial outflow solution is preserved. {Right}: Relative differences between
    the linear perturbed analytical solution and the convergence numerical solution in density, radial velocity and angular velocity, top to
    bottom. {  The top two pannels giving curves for three values of the angle, $\theta=0^{\circ}$, $\theta=54.74^{\circ}$ and $\theta=90^{\circ}$,
    purple, blue and green lines respectively. Given the symmetry conditions, the angular velocity vanishes for both $\theta=0^{\circ}$  and
    $\theta=90^{\circ}$, the bottom panel shows angular velocity difference curves for  $\theta=0^{\circ}$, $\theta=54.74^{\circ}$ and $\theta=90^{\circ}$,
    purple, blue and green curves respectively.}
  }
\end{figure*}

Given that the radial accretion model of eqs. (13) and (14) is an exact solution of the conservation equations of the problem, the excellent
agreement of the numerical solution to these equations is not a validation of the analytical model, but of the numerical procedure implemented.
Indeed, we are not aware of any other exact spherical accretion solution for the case of $\gamma = 5/3$ which can be used to test numerical
codes, as numerical experiments testing against the classical Bondi solution typically fail on reaching exactly $\gamma = 5/3$, e.g.
Waters et al. (2020).

Lastly, we go to a 2D numerical setup under cylindrical symmetry using 256 radial cells and 128 angular ones, to explore the non-spherical
perturbations to the exact radial solution, the results of section 3. Again taking a sub-sonic case, this time  $\varrho^{2/3}=0.56$, the initial
conditions were taken from eqs. (54), (55), (56) and (53). In this case we used $\epsilon
\varrho_{1}=0.0005$, so as to remain close to the validity of the linear analysis at small radii, the plus sign was taken in eq.(53), leading to
the equatorial infall/radial outflow solution. Results are given in figure 8, where the left panel shows the convergence numerical solution in
terms of flow lines and density field. This final condition is visibly different from the initial conditions used, in that the stagnation point
in the radial velocity along the polar direction has shifted outwards a little, but otherwise, the general structure of the solution is preserved,
e.g compare to Fig. 6, left panel. Some adjustments on approaching the stagnation point and beyond are to be expected, as when the radial velocity
goes to zero, the linear approximation leading to eqs.(53)-(56) breaks down. Crucially, the overall character of the solution remains unchanged,
validating the infall/outflow character of the perturbed analytic solution.

In more detail, the right panel of figure 8 gives the logarithms of the relative
differences between the analytic solution of eqs.(53)-(56) and the final numerical convergence solutions, once the spherical component, the solution
of eqs. (13), (14) has been subtracted, so as to yield a comparison only of the non-spherical components. The right panel gives the quantities mentioned
for three values of the angle, $\theta =0^{\circ}$, $\theta=54.74^{\circ}$ and $\theta=90^{\circ}$, purple, blue and green lines respectively, for the
two top panels. As given the symmetry conditions the angular velocity vanishes for both  $\theta =0^{\circ}$ and $\theta=90^{\circ}$, the three angles
chosen for the bottom panel where  $\theta =34.74^{\circ}$, $\theta=54.74^{\circ}$ and $\theta=74.74^{\circ}$. $\theta=54.74^{\circ}$ is the critical angle
where the linear analytic solution yields $\mathcal{V}=0 $ for large radii.

We can see that for most of the density and velocity radial profiles shown, the numerical convergence solution differs from the input
approximate one by factors of less than order one. Although for some angles and some radii this is not the case and such differences
become larger, as can be seen from the overall flow pattern resulting in the left panel of figure 8, the general equatorial infall and
polar outflow character of the steady state solution is maintained. Thus, the changes are quantitative but not
qualitative, the general structure is preserved even after the end of the validity regime of the linear approximation.

We note that the cases presented here cover a very small region of the parameter space to be tested numerically; deviations from  $\gamma = 5/3$,
temporal perturbations, dynamical cases exploring the convergence to the analytic cases from various initial conditions, more general angular
variations, and various combinations of the above need to be explored in order to better understand the full richness of the solutions
presented in the previous sections.

Here we present only a brief preliminary series of tests showing the numerical stability of the exact
spherical solution, and that even after crossing the validity regimes of the linear perturbation study presented in section 3, the perturbed
solution converges to a steady state preserving the qualitative features of the perturbative solution described in section 3. A subsequent
series of papers will explore the very large number of clearly relevant variants in detail.

 }

\section{Conclusions}

We have presented a new, exact steady-state spherically-symmetric zero angular momentum hydrodynamic accretion model for $\gamma=5/3$
in a Keplerian potential. It differs from classical Bondi accretion in that the condition at infinity has been changed from $\rho \to \rho_{\infty}$
to $\rho \to 0$, which allows simple power law solutions for the density and the infall velocity in the spherically symmetric case.
These solutions are characterised by having a Mach number,
$\mathcal{M}$, which is constant for all radii, at supersonic values towards the cold, empty limit, and at subsonic ones towards the hot,
dense hydrostatic equilibrium one. A maximum accretion rate appears precisely at $\mathcal{M}=1$. It is interesting that in the limit
as $R \to 0$, the $\gamma=5/3$ Bondi solution actually converges to the new solution we present here. This new solution is stable
to small temporal deviations, and also to small changes in the adiabatic index about $\gamma=5/3$.

Having an exact solution in closed form for the spherical accretion problem allows its use to explore deviations from sphericity with the
polar angle in velocity and density fields and the inclusion of angular momentum, through an analytic perturbative analysis. This yields an ample
spectrum of solutions given by Legendre polynomials for the density field, through which the velocity fields follow.

As a first exploration we show simple bipolar configurations having equatorial infall and polar outflows, where interestingly, one branch
can naturally yield polar outflows for radii larger than a critical value, still within the consistency region of the perturbative analysis
performed on the density field. Although the formal validity limit of the perturbative solution in the velocity field is reached before
said point, it is interesting that even beyond this point the analytical solution is qualitatively consistent with recent independent numerical
experiments. For the same solution, a second branch alters the spherical symmetry solution only at small radii, and does so by redirecting
the large-scale spherical infall into a wedge along the equatorial plane, also well within the validity regime in the treatment
of the density field. This effect is minor within the validity regime of the velocity approximation, and grows to becomes dominant
on approaching the velocity validity threshold, suggesting the possibility of an interesting mechanism for super-Eddington accretion.

{  A first series of numerical experiments confirms the temporal stability of the new spherically symmetric solution. Simulations
  in 2D allowing for polar angle deviations from sphericity deviate quantitatively from the simple analytic linear angular perturbation
  solutions of section 3 which consider only one term in the Legendre series, while maintaining the overall equatorial accretion and
  radial outflow structure of such solutions.}

Given the infinite range of behaviours which can be modelled using the Legendre polynomials solving eq.(37), it is reasonable
to expect the framework presented might be of interest in a range of astrophysical settings.

\section*{Acknowledgements}

The authors acknowledge the careful reading and constructive criticism of an anonymous referee as valuable towards reaching a more complete
final version. The authors also thank Tom\'{a}\v{s} Pl\v{s}ek for kindly making the data of Pl\v{s}ek et al. (2022) available in electronic form. The authors
wish to thank Nissim Fraija for helpful discussions during the preparation of this manuscript. Xavier Hernandez acknowledges financial assistance
from UNAM DGAPA grant IN106220 and CONACYT. L. Nasser gratefully acknowledges the support from the NSF award PHY - 2110425.

\appendix

\section {Comparison of spherical solution to observed accretion profiles from Pl\v{s}ek et al. (2022)}

{ 
We here show the remaining 18 density accretion profiles from Pl\v{s}ek et al. (2022) along with $R^{-3/2}$ fits to the data, the expected density scaling
from the new solution of eqs. (13) and (14). As explained in section (2), the first point in each profile is reported as having a 100\% error,
as described by the above authors this innermost point is a model extrapolation rather than a direct measurement, and hence
was excluded from the fits. In about half the cases, such as NGC 4261, NGC 4552 or the two galaxies included in section 2, the fits are very good, the
radial scaling predicted by the new model presented quite accurately reproduces the data. In about a third of the cases (e.g. NGC 4472, NGC 4696 or
NGC 5813) the observed profiles still show the predicted scaling over a substantial intermediary radial range, but show also inner and/or outer flattening
with respect to the single power-law expectations of the solution of eqs. (13) and (14), signaling a validity regime for the solution presented. Towards
the inner regions this could be due to the effects of AGN activity, as described in e.g. Runge \& Walker (2021), whilst towards the outer regions the presence
of radio lobes or a regime change from the inner accretion profile to an outer unperturbed medium could explain the deviations mentioned, amongst many other
effects not considered in the very simple spherically-symmetric accretion model presented. Finally, in three cases, NGC 507, NGC 5044 and NGC 6166, the fit
presented is not a good match to the data, which show not only inner and/or outer flattening with respect to a single power law fit, but also a power law
scaling in the intermediary regions incompatible with the $R^{-3/2}$ of the fits.

{  Further insight into the possible causes determining the existence or otherwise of extended radial ranges compatible with the $R^{-3/2}$ accretion
density profiles can be gained by considering correlations with other observed properties for these systems. As a first such exploration, we can
look at the presence or absence of $H\alpha+[N_{II}]$ line emitting gas extended over central regions larger than 2kpc. This can be an indication of
the presence of thermally unstable atmospheres (Pl\v{s}ek et al. 2022) and hence more complex physical conditions falling outside of the simple
model presented in section 2. Half of the sample, 10 systems, are reported by Pl\v{s}ek et al. (2022) as presenting such extended emission,
and half as not presenting this feature. We take a first order appraisal of the degree to which the $R^{-3/2}$ density profile represents the data,
where fits are classified poor (P) when at least two data points (excluding the innermost) lie at a horizontal distance of more than 1.5$\sigma$ from
the  $R^{-3/2}$ density profile and as good (G) otherwise. Also, from Pl\v{s}ek et al. (2022) we know which systems show $H\alpha+[N_{II}]$ line emitting gas
over an extent larger than 2kpc, (E), and which do not (N). The full list of twenty galaxies is now: IC 4296 (E,G), NGC 507 (N,P), NGC 708 (E,P),
NGC 1316 (E,G), NGC 1399 (N,G), NGC 1407 (N,G), NGC 1600 (N,G), NGC 4261 (N,G), NGC 4374 (N,G), NGC 4472 (N,P), NGC 4486 (E,P), NGC 4552 (E,G), NGC 4636 (N,G),
NGC 4649 (N,G),  NGC 4696 (E,P), NGC 4778 (N,P), NGC 5044 (E,P), NGC 5813 (E,G), NGC 5846 (E,G) and NGC 6166 (E,P), where we note for each system the two
above classifications. It is interesting that cases presenting extended $H\alpha+[N_{II}]$ are almost twice as likely to be classed as poor fits (5/10), than cases
not presenting such extended emission (3/10).

This is clearly only a first attempt at correlating observational indicators of perturbed physical conditions in these systems with the degree to which the
density profile of the new solution presented fits the observed accretion profiles, the small present sample of high quality observations does not warrant
any further refinements of these explorations at this point, which also fall beyond the intent of our present study.}

The figure captions gives the inferred values of the Bondi radii for each galaxy, also from Pl\v{s}ek et al. (2022), in all cases
values much smaller than the middle range of the radial values shown, often smaller than the lower radial range. Hence, classical Bondi fits to the accretion
density profiles shown would look very close to horizontal lines over most of the radial ranges shown in the log-log plots given. Therefore, once the
inferred Bondi radii are fixed for all galaxies, even for the cases where the new proposed fit is poor, this remains a much better fit than a classical Bondi
profile.
}

\begin{figure*}
 \includegraphics[height=5.1cm,width=5.5cm]{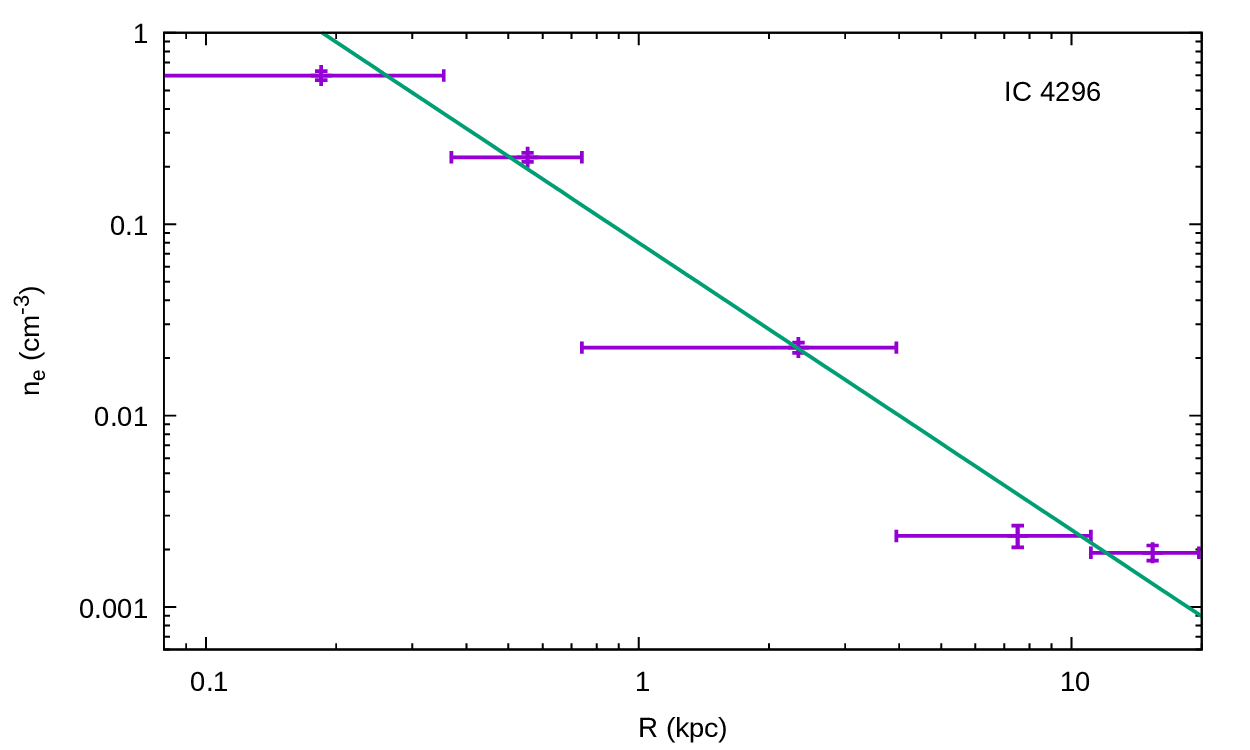} \includegraphics[height=5.1cm,  width=5.5cm ]{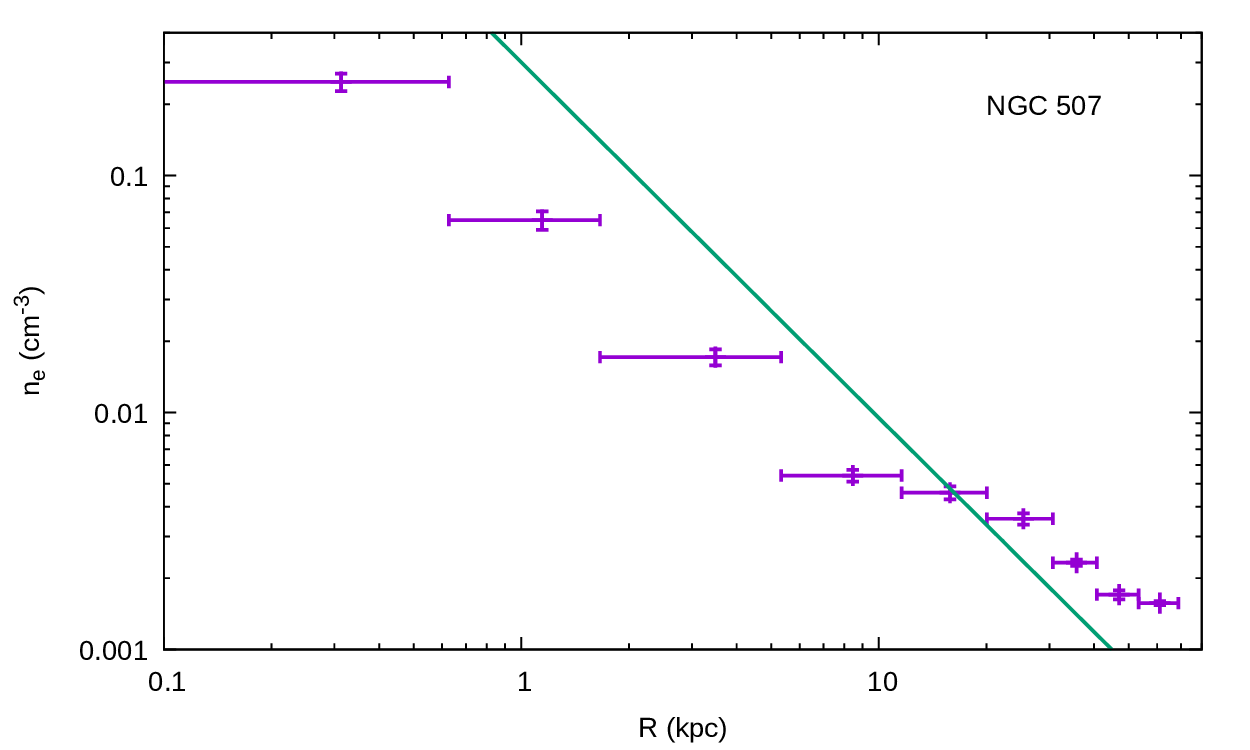} \includegraphics[height=5.1cm,  width=5.5cm ]{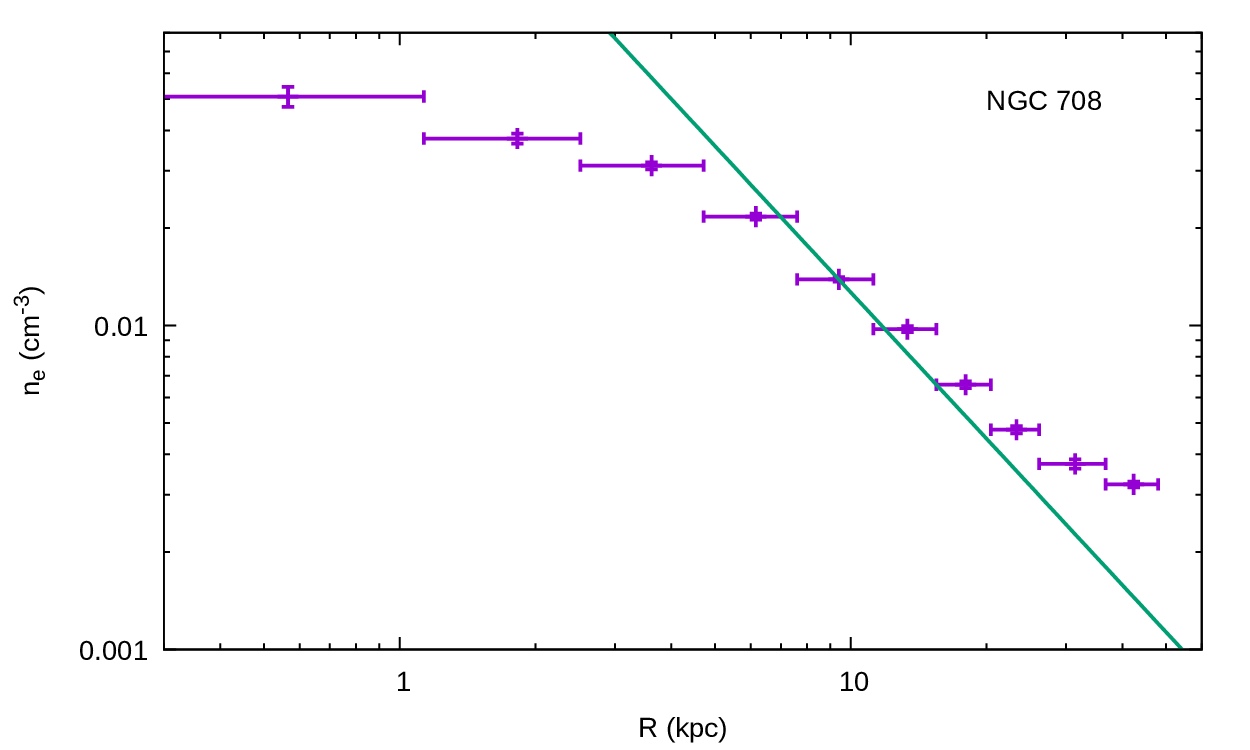} 
 \includegraphics[height=5.1cm, width=5.5cm ]{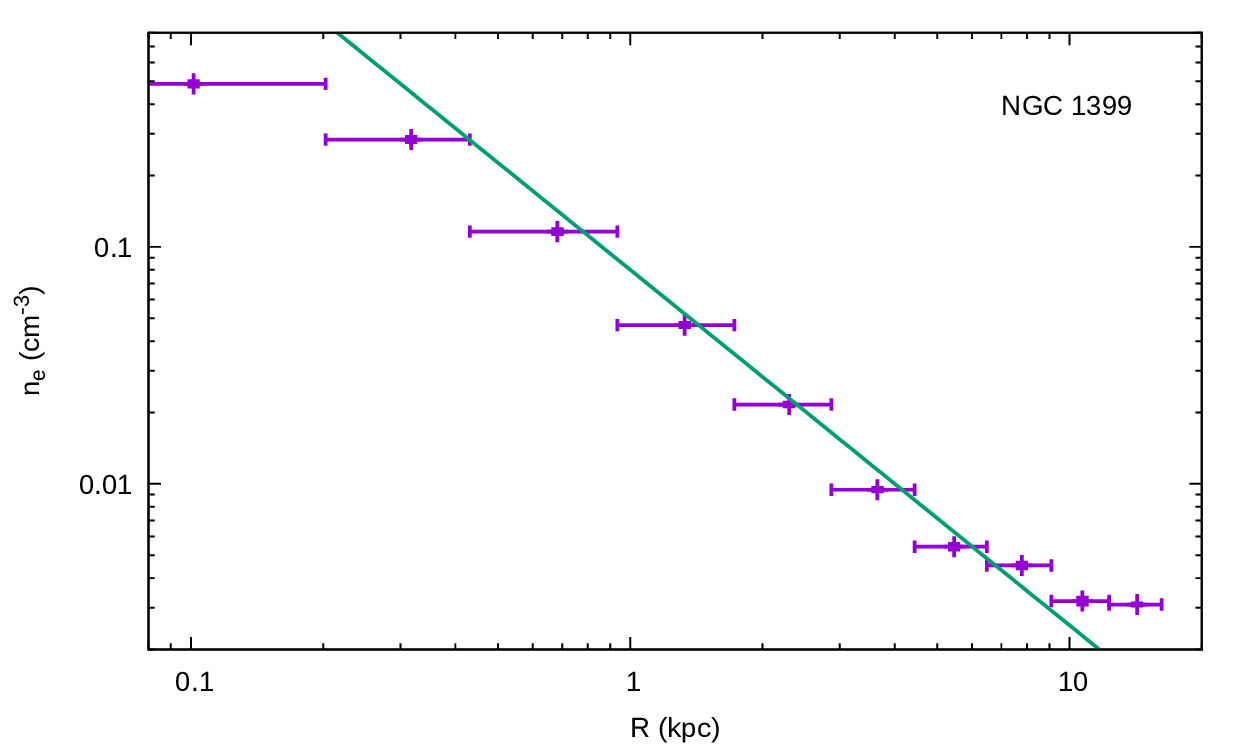} \includegraphics[height=5.1cm, width=5.5cm ]{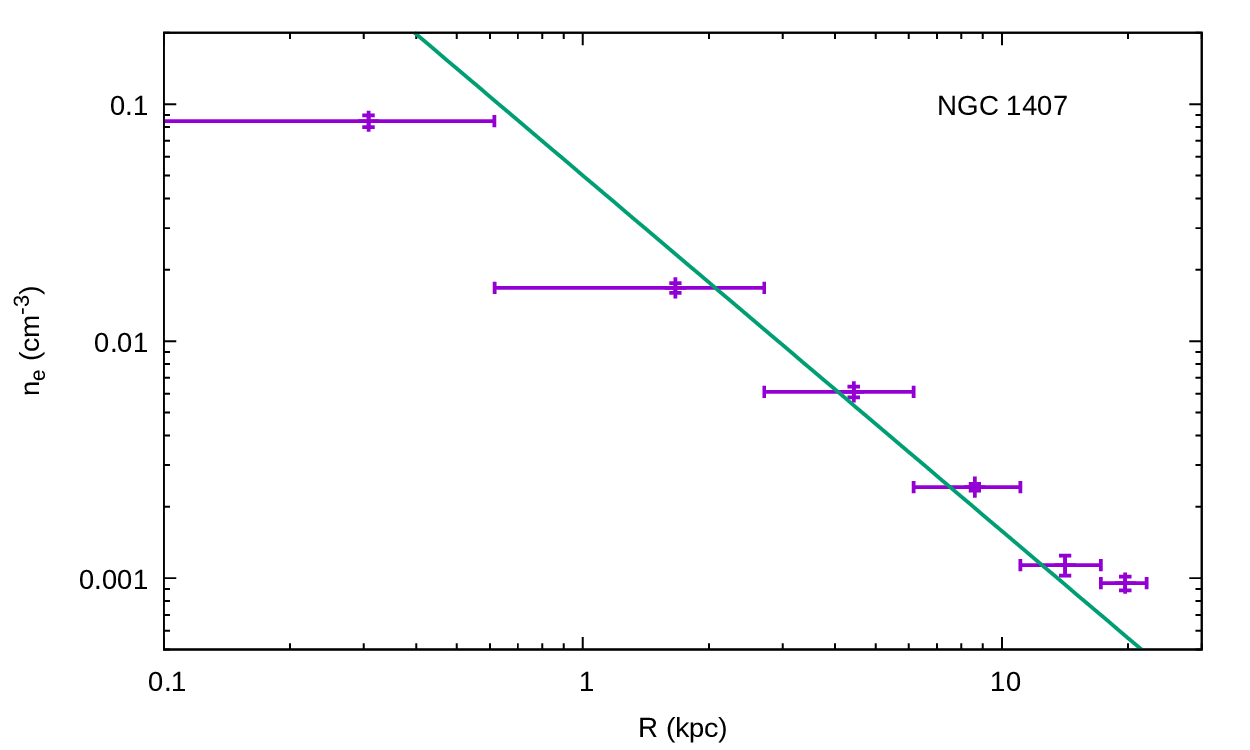} \includegraphics[height=5.1cm,width=5.5cm]{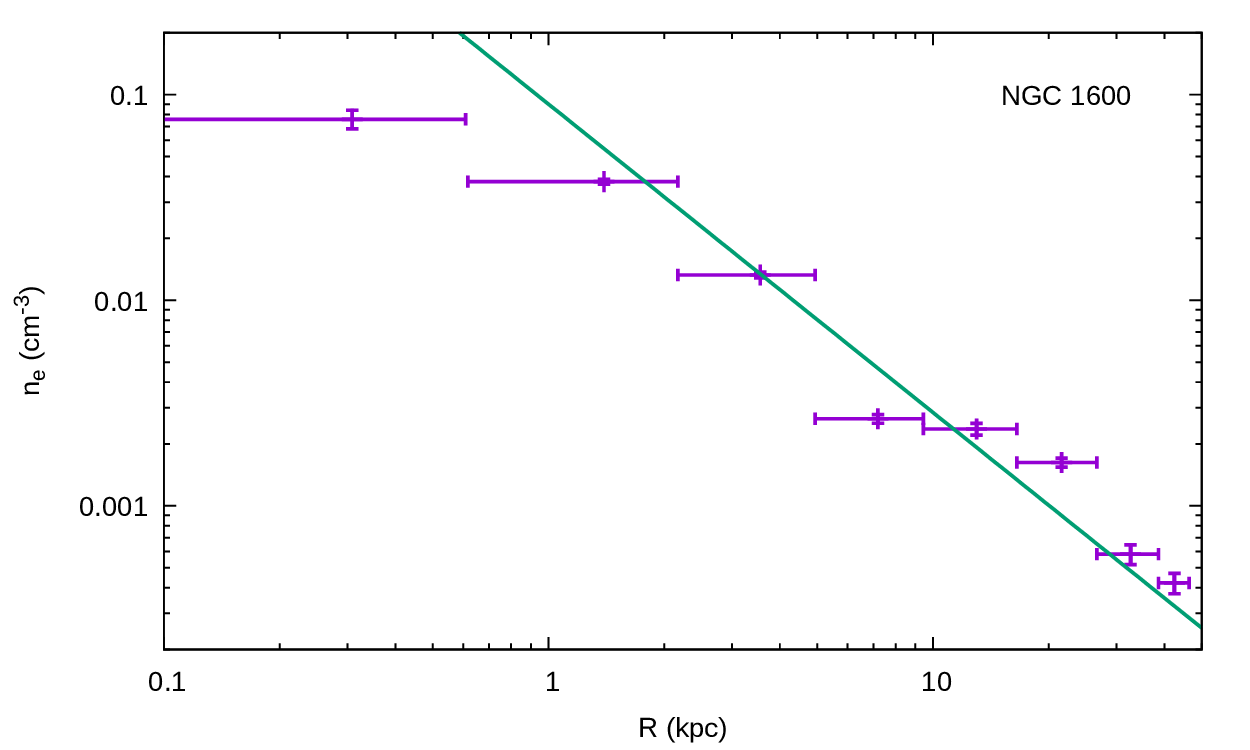}
 \includegraphics[height=5.1cm, width=5.5cm ]{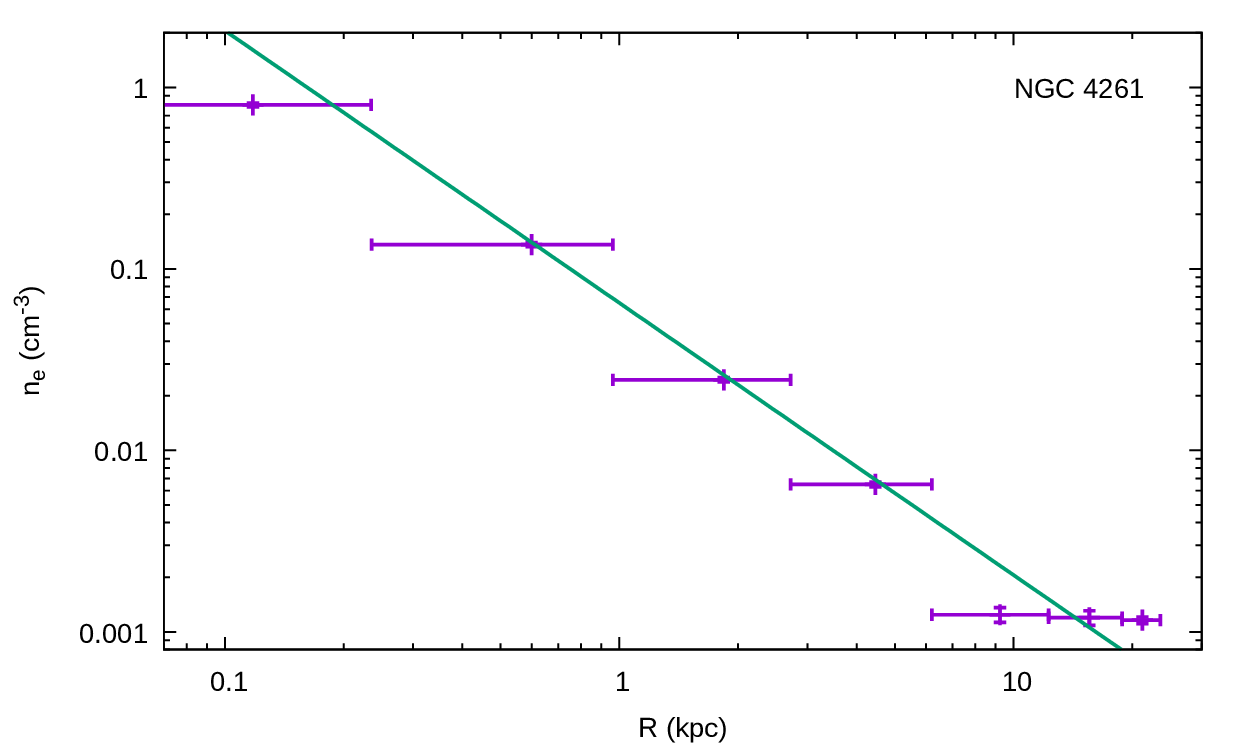} \includegraphics[height=5.1cm, width=5.5cm ]{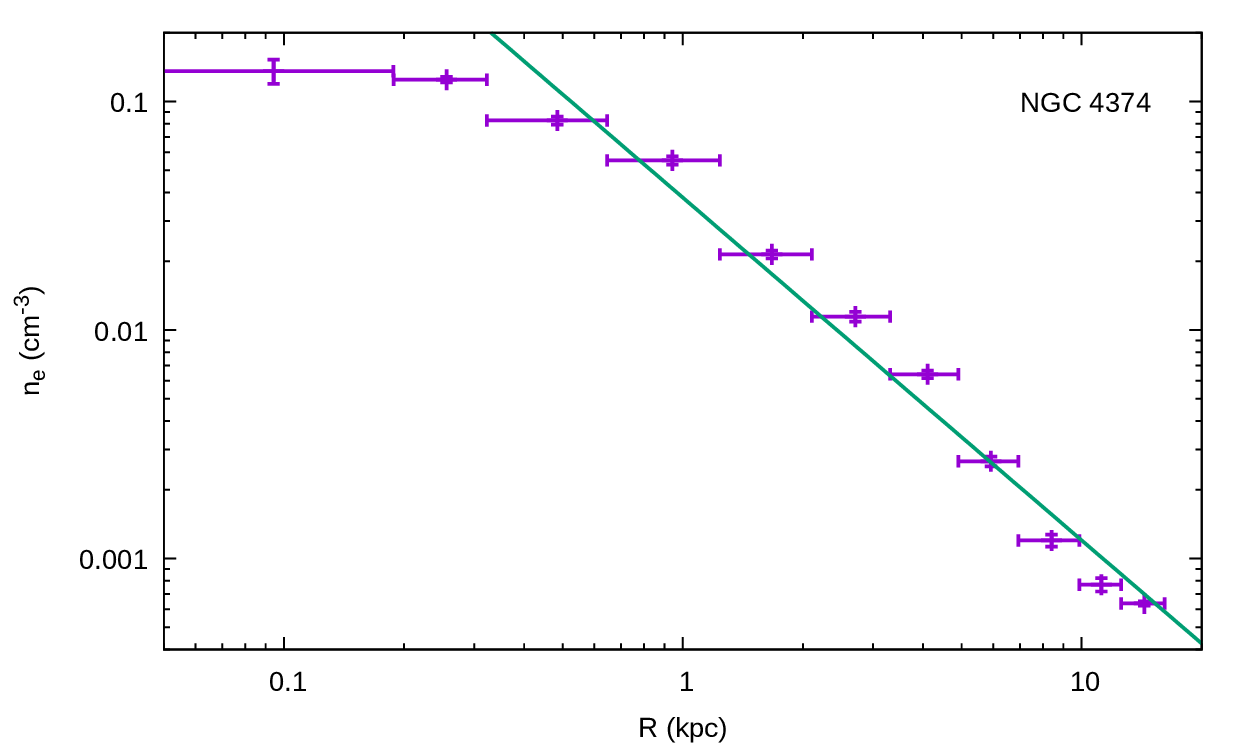} \includegraphics[height=5.1cm,width=5.5cm]{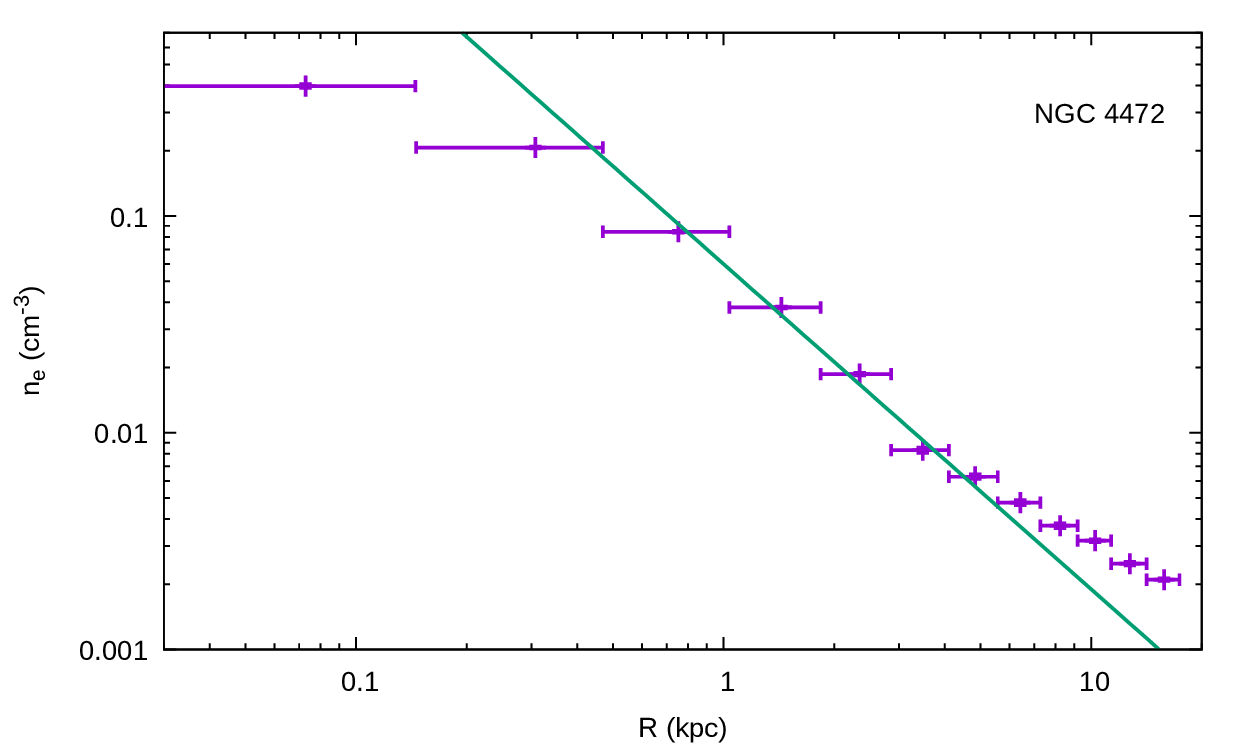}
 \includegraphics[height=5.1cm, width=5.5cm ]{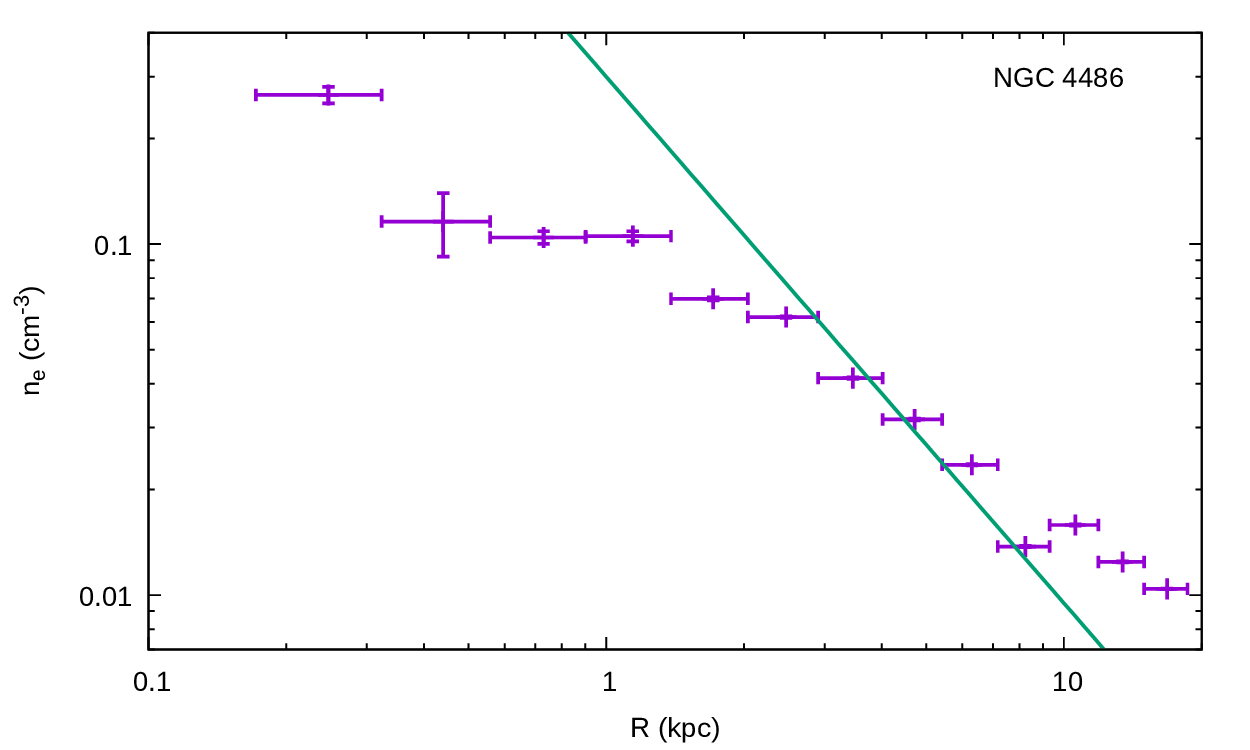} \includegraphics[height=5.1cm, width=5.5cm ]{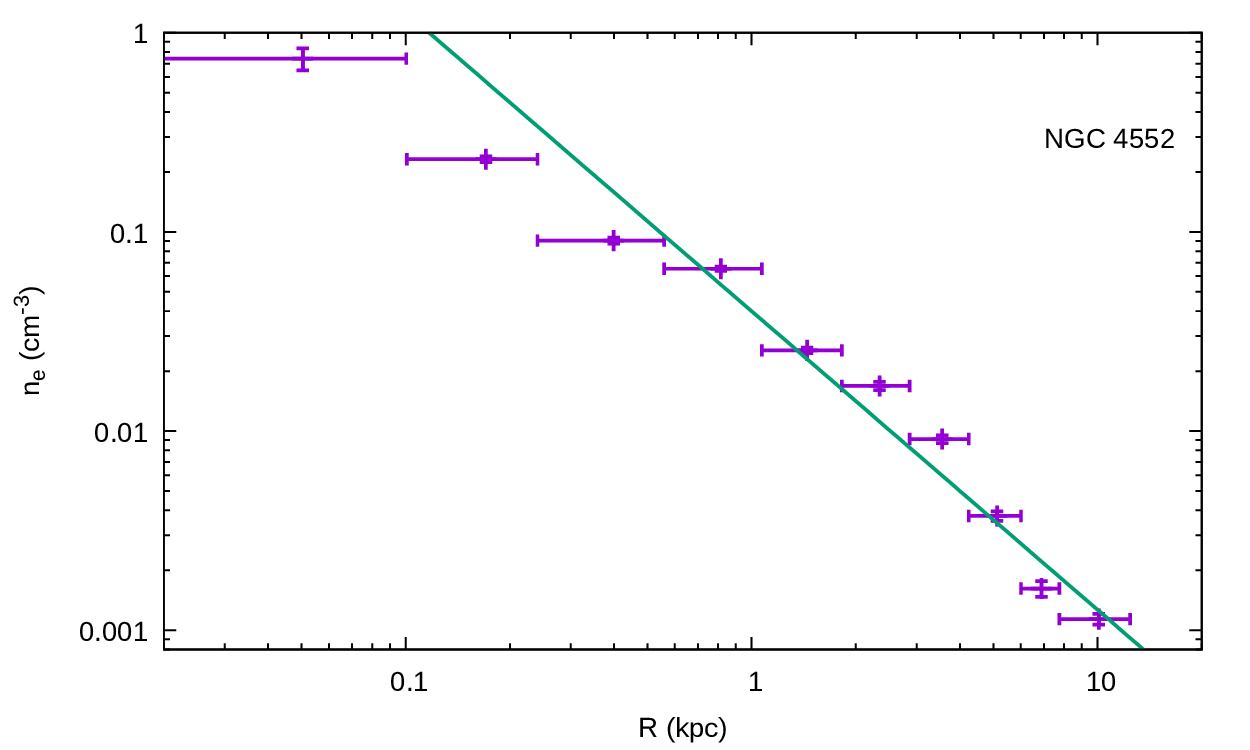} \includegraphics[height=5.1cm,width=5.5cm]{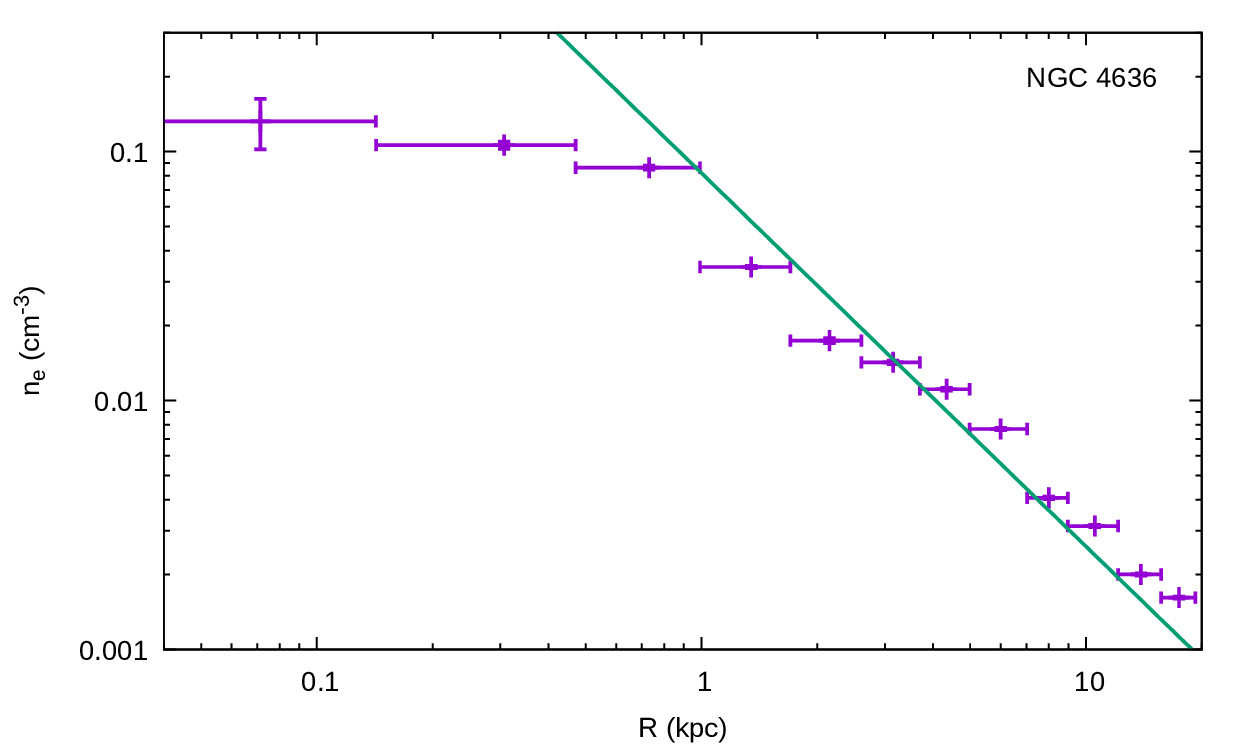}

  \caption{Inferred electron density profiles about the central SMBHs of 12 galaxies as presented by Pl\v{s}ek et al. (2022), points with error bars.
    The solid lines give $R^{-3/2}$ fits, as expected by the new solution of eqs. (13) and (14). The Bondi radii of the galaxies shown are of
    $63^{+13}_{-11}$ pc for IC 4296, $87^{+24}_{-20}$ pc for NGC 507, $16^{+3}_{-2}$ pc for NGC 708,  $37^{+15}_{-17}$ pc for NGC 1399,
    $170^{+30}_{-20}$ pc for NGC 1407, $590\pm 50$ pc for NGC 1600, $80^{+19}_{-12}$ pc for NGC 4261, $42 \pm 4$ pc for NGC 4374, $100^{+18}_{-10}$ pc for NGC 4472, 
    $270 \pm30$ pc for NGC 4486, $15^{+3}_{-2}$ pc for NGC 4552 and $37 \pm 5$ pc for NGC 4636, as reported by Pl\v{s}ek et al. (2022). Classical Bondi fits
    would look very close to horizontal over the radial range shown, given the constraint of the inferred values for the Bondi radii.}
\end{figure*}

\begin{figure*}
  \includegraphics[height=5.1cm, width=5.5cm ]{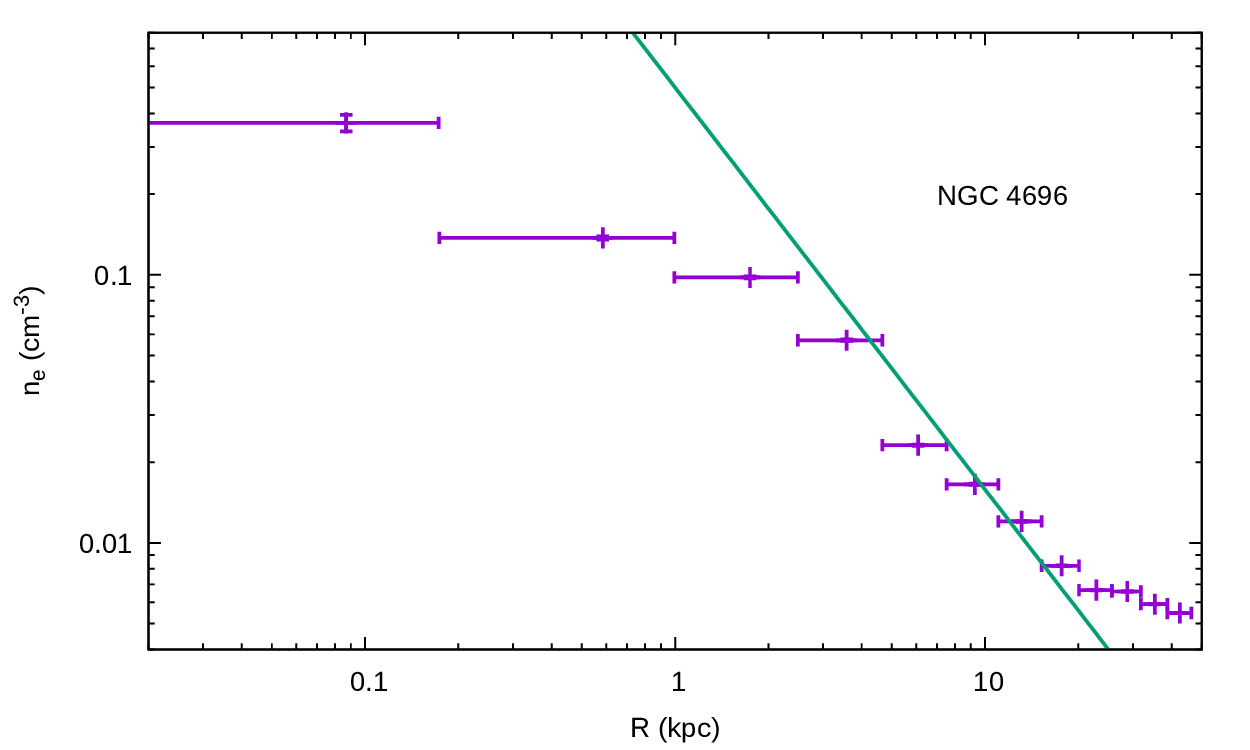} \includegraphics[height=5.1cm,width=5.5cm]{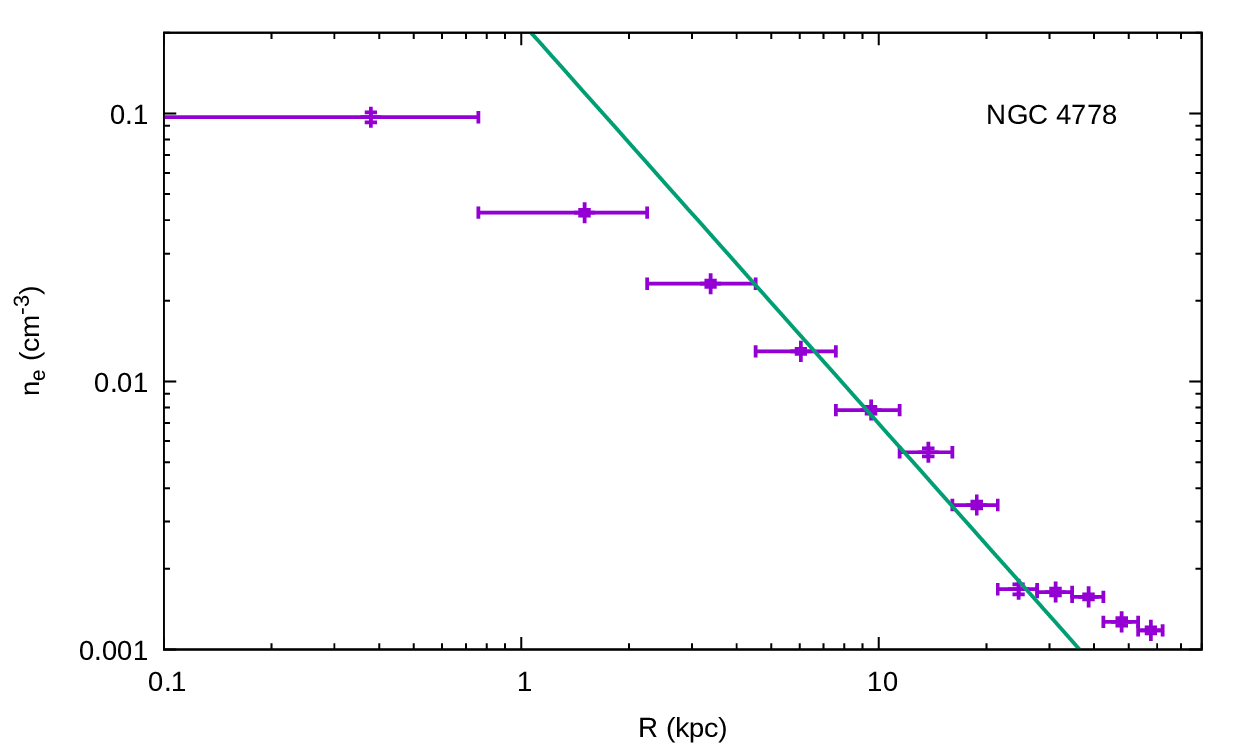} \includegraphics[height=5.1cm, width=5.5cm ]{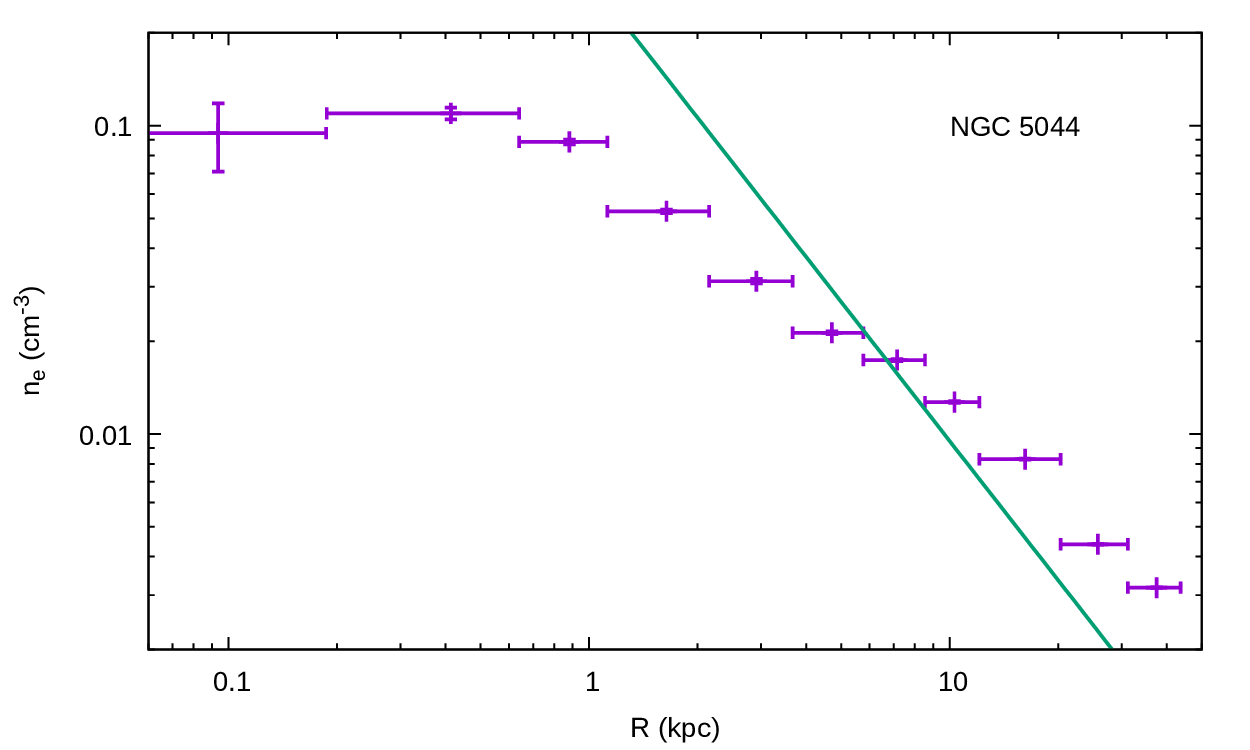}
  \includegraphics[height=5.1cm, width=5.5cm ]{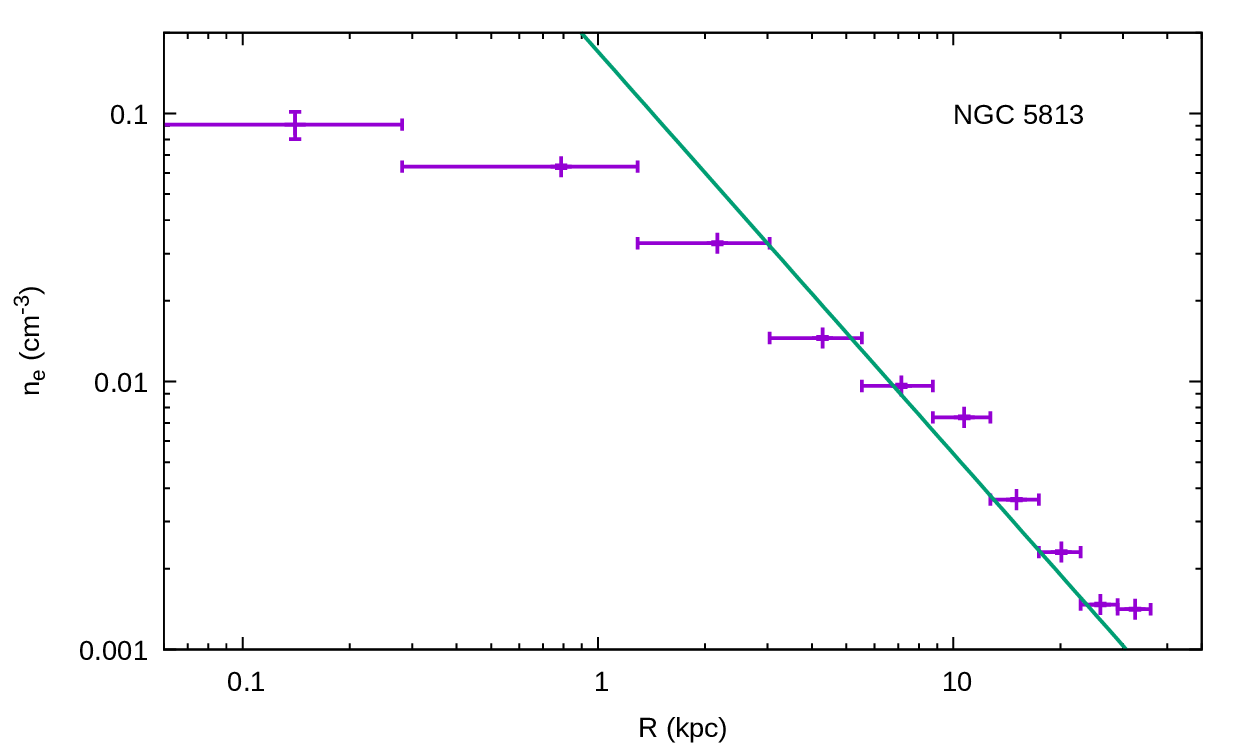} \includegraphics[height=5.1cm,width=5.5cm]{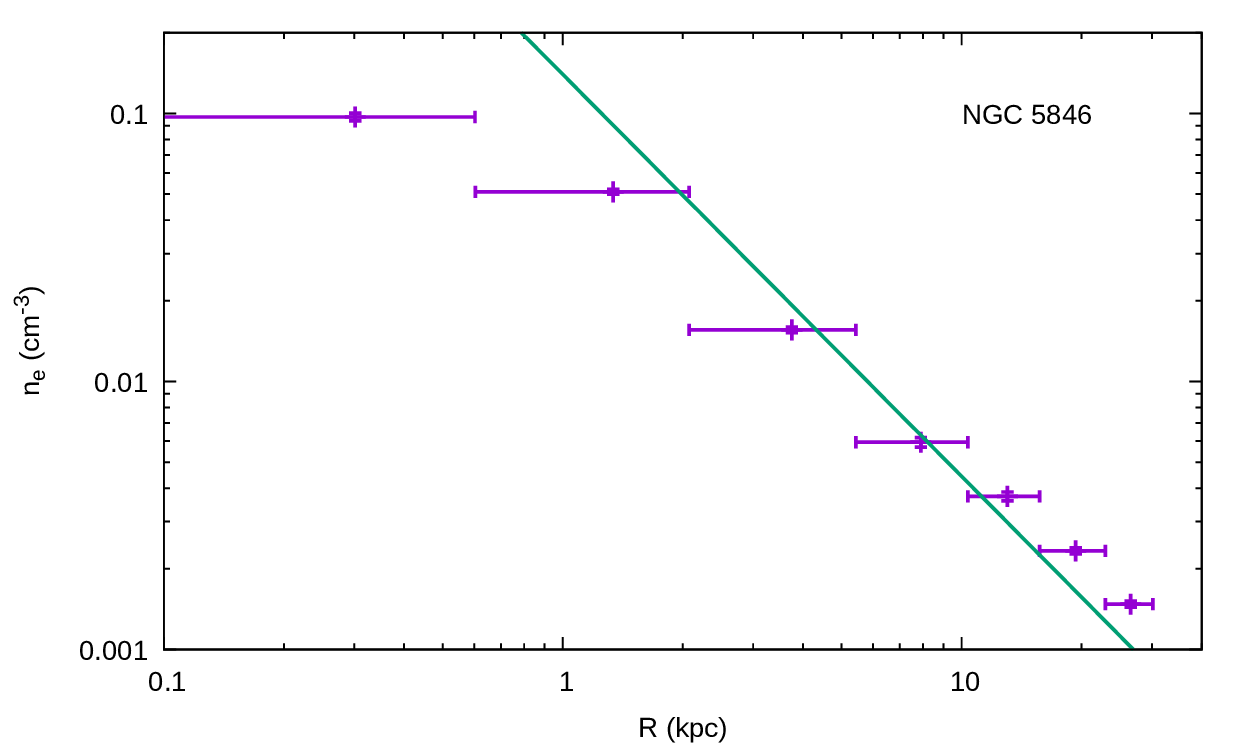} \includegraphics[height=5.1cm, width=5.5cm ]{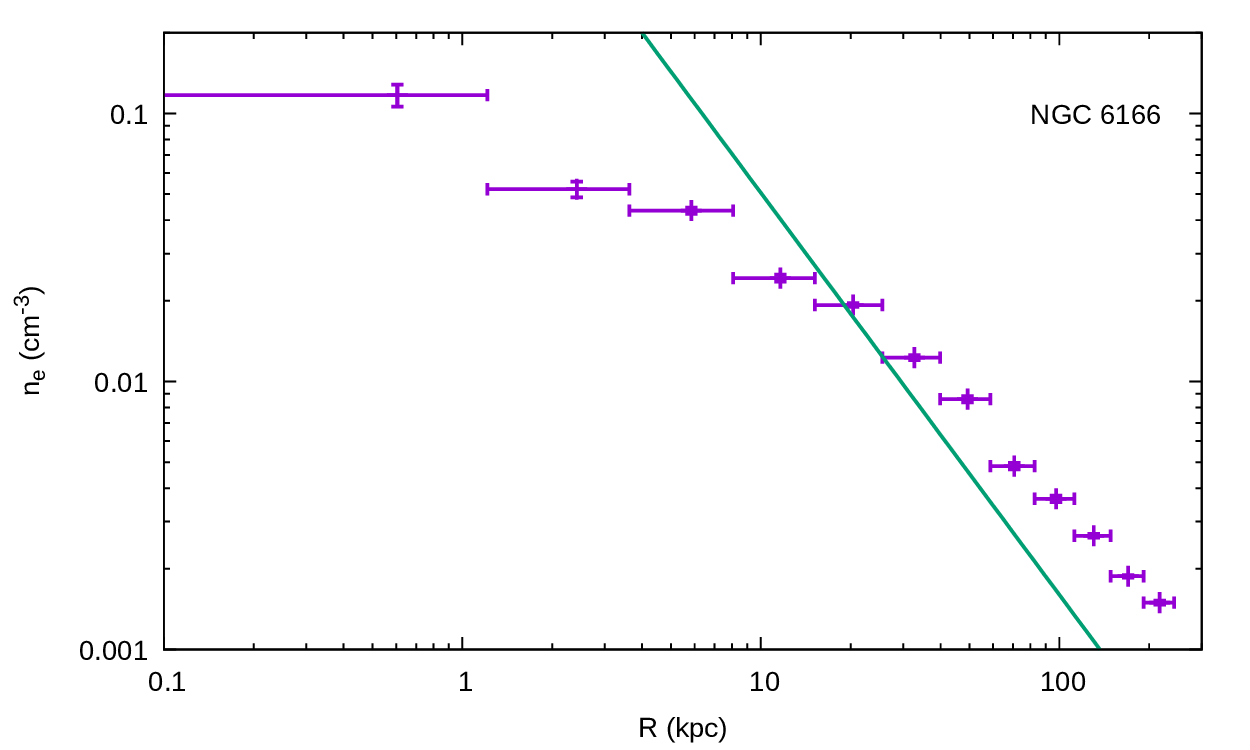} 
  \caption{Inferred electron density profiles about the central SMBHs of the 6 remaining galaxies of the 20 as presented by Pl\v{s}ek et al. (2022), points with
    error bars. The solid lines give $R^{-3/2}$ fits, as expected by the new solution of eqs. (13) and (14). The Bondi radii of the galaxies shown are of
    $34^{+7}_{-6}$ pc for NGC 4696, $37^{+8}_{-6}$ pc for NGC 4778, $9.9^{+5.3}_{-3.2}$ pc for NGC 5044, $30 \pm 4$ pc for NGC 5813, $48^{+7}_{-6}$ pc for NGC 5846 and 
    $44^{+13}_{-10}$ pc for NGC 6166, as reported by Pl\v{s}ek et al. (2022). Classical Bondi fits would look very close to horizontal over the radial range shown, given
    the constraint of the inferred values for the Bondi radii}
\end{figure*}

\section{Appendix: Variations of the spherical solution about $\gamma=5/3$.}

We now explore small variations about $\gamma=5/3$ to test for the stability of the solution in eqs. (13)-(15) to departures from
the adiabatic value for which said solution was developed. We now take $\gamma=5/3+\mu$, where we shall assume $\mu <<1$.

The spherically symmetric solution will be modified to:

\begin{equation}
\mathcal{V}=\mathcal{V}_{0}R^{\nu-1/2},
\end{equation}

\begin{equation}
\varrho=\varrho_{0}R^{-\nu-3/2}.
\end{equation}

For the above solution it is clear that the mass conservation equation, eq.(11), will still be satisfied identically, as the power law
departures for $\gamma \neq 5/3$ in $\mathcal{V}$ and $\varrho$ will cancel, leaving the product  $\mathcal{V}\varrho$ unchanged.

The momentum conservation equation however, eq.(12), will no longer be satisfied identically, and is only satisfied to leading order in the
deviations with respect to the $\gamma=5/3$ solution, under the assumption of  $\mu <<1$ and  $\nu <<1$. Equation (12) becomes:

\begin{equation}
(\nu-1/2) \mathcal{V}_{0}^{2} R^{2\nu-2}=(\nu+3/2)\varrho_{0}^{\mu+2/3}R^{-2\nu/3-3\mu/2-2}-R^{-2}
\end{equation}

We shall now introduce a series expansion for $R^{A+\epsilon} = R^{A}+\epsilon R^{A} \ln(R)+\mathcal{O}\epsilon^{2}$ to develop the radial
powers in the above equation, yielding:

\begin{equation}
\begin{split}
  (\nu-1/2) \mathcal{V}_{0}^{2}R^{-2}+2\nu(\nu-1/2) \mathcal{V}_{0}^{2}R^{-2}\ln(R)=\\
  (\nu+3/2)\varrho_{0}^{\mu+2/3}R^{-2}\\-(2\nu/3+3\mu/4)(\nu+3/2)\varrho_{0}^{\mu+2/3}R^{-2}\ln(R).
\end{split}
\end{equation}

\noindent The terms without $\ln(R)$ in the above equation yield:

\begin{equation}
(\nu-1/2) \mathcal{V}_{0}^{2}=(\nu+3/2)\varrho_{0}^{\mu+2/3}-1,
\end{equation}

\noindent while the terms containing $\ln{R}$ yield to first order in $\mu$ and $\nu$:

\begin{equation}
\nu \mathcal{V}_{0}^{2}=(\nu+9\mu/4)\varrho_{0}^{\mu+2/3}.
\end{equation}

\noindent Equations (C5) and (C6) can now be used to obtain the departure in the power laws of the solution of eqs. (C1), (C2) to our
original spherical solution, $\nu$, as a function of the departure from $\gamma=5/3$, $\mu$, and the parameter of the spherical solution,
$\varrho_{0}$ :

\begin{equation}
\nu=\frac{9 \mu \varrho_{0}^{2/3+\mu}}{8(1-2\varrho_{0}^{2/3+\mu})}.
\end{equation}

\noindent Notice that as the departure from $\gamma=5/3$ goes to zero, $\mu \to 0$, the departure from the original solution also goes to zero,
as is also the case towards the empty free-fall solution of $\varrho=0$, where in the absence of hydrodynamical effects the free-fall solution
is recovered, regardless of the value of the adiabatic index. Condition (15) is modified slightly in this case to yield:

\begin{equation}
\mathcal{V}_{0}^{2}=2-3\varrho_{0}^{2/3+\mu}.  
\end{equation}

\noindent With the exception of a divergence in $\nu$ close to the $\mathcal{M}=1$ value of $\varrho^{2/3}_{0}=1/2$, small deviations from
$\gamma=5/3$ result in small deviations in the power laws of the original spherically symmetric solution. {  Hence the power law character of
the solution in eqs. (13) and (14) is generally preserved with only small variations when changing $\gamma$ slightly away from the idealised
$5/3$ value}. As expected, $\mu \to 0$ recovers the original solution for $\gamma =5/3$. We see that the spherically symmetric solution of
eqs(13) and (14) is robust to small changes in the adiabatic index.

{ 
We can now follow the same procedure of section 2 to calculate the Mach number of the spherically-symmetric solution, but this time using the
solutions for slight deviations from $\gamma=5/3 \to \gamma=5/3+\mu$, equations (C1) and (C2) instead of equations (13) and (14). Assuming both
$\mu<<1$ and $\nu <<1$, to first order in these variables we obtain:

\begin{equation}
\mathcal{M}=\mathcal{V}_{0}\varrho_{0}^{-(1/3+\mu/2)} R^{3\mu/4+4\nu/3},
\end{equation}

\noindent using the result of (C7) we obtain to first order:

\begin{equation}
\mathcal{M}=\mathcal{V}_{0}\varrho_{0}^{-(1/3+\mu/2)} R^{3\mu/(4-8\varrho_{0}^{2/3})}.
\end{equation}

\noindent Clearly, when $\gamma \to 5/3$ both $\mu \to 0$ and $\nu$ $\to 0$, and the result of eq.(20) for the constant $\mathcal{M}$ of the solution in
section 3 is recovered, showing the consistency of the above analysis.

The above result is interesting because it shows that the radial constancy of the Mach number obtained for $\gamma=5/3$ is lost when considering
even perturbative deviations in the adiabatic index away from $5/3$, while retaining the  $\rho \to 0$ as $r \to \infty$ condition. Whilst the
power law character of the density and velocity fields in the spherical solution is robust to small changes in $\gamma$ (provided one is not close
to $\varrho_{0}^{2/3}=1/2$), the lack of a sonic radius is not. As shown in eq.(C10), even small deviations away from  $\gamma=5/3$ result in a
radial profile in the Mach number of the solution which is no longer flat. This profile can be slightly increasing, or slightly decreasing with
radius, depending on the signs of $1-2\varrho_{0}^{2/3}$ and of $\mu$. When the two signs are different, an outer sub-sonic region results, which
transitions to an inner super-sonic one at a sonic radius, much like the classical Bondi solution. However, when both signs are equal, the revers
ensues, an outer $\mathcal{M}>1$ region, a sonic radius and an inner $\mathcal{M}<1$ one. This last feature does not imply any inward deceleration,
the velocity flow in all cases remains that of eq.(C1), which for the assumption of $\mu<<1$ and $\nu<<1$, under which the results of this appendix
were derived, implies a flow which always accelerates inwards, much like the one of the $\gamma=5/3$ solution of section 3. The transition from
$\mathcal{M}>1$ at large radii to the opposite at small ones mentioned above, arises merely from the fact that the sound speed increases at a faster
rate towards small radii than the flow velocity.

For the exact solution to the conservation equations with $\gamma=5/3$ in the spherical case of eqs.(13) and (14), the radial power law scalings of
the local sound speed and the infall velocities result equal, yielding the constant Mach number discussed in section 3. However, for even perturbative
departures from $\gamma =5/3$, both radial dependences discussed above change in slightly different manners. As seen in eq.(C10), in this case
$\mathcal{M}$ will no longer be constant and a sonic radius will appear. Still, the radial variation in $\mathcal{M}$ will remain small, provided $\mu$
is small and $\varrho_{0}^{2/3}$ is not close to $1/2$. If either of these last two conditions are not met, the analysis in this appendix will no longer
be valid, but from the developments leading to eq.(C7) we see that when $\varrho_{0}^{2/3}$ approaches $1/2$, even for small $\mu$, large deviations will
appear in the velocity and density fields of the solution, and hence also in the radial profile of $\mathcal{M}$, when moving away from $\gamma=5/3$.
Thus, the constant $\mathcal{M}$ of the eqs. (13) and (14) solution is only a feature of the idealised $\gamma$ exactly equal to $5/3$ case. Still, within
the validity regime of the above perturbative analysis, the radial profile of  $\mathcal{M}$ remains in general much flatter than for the case of
the classical Bondi solution.
}

\end{document}